\documentclass[a4paper,11pt]{article}
\usepackage{aaskaiid}
\usepackage{hyperref}
\usepackage{orcidlink}

\title{Revisiting the Unknown Unknowns in Transient Astrophysics with the SKA Telescopes}
\ShortTitle{Unknown unknowns}

\author[1]{Alex Andersson\orcidlink{0000-0003-2734-1895}}
\author[2]{Manisha Caleb\orcidlink{0000-0002-4079-4648}}
\author[3,7]{Joseph Callingham}
\author[4]{Adam Deller\orcidlink{0000-0001-9434-3837}}
\author[2]{Dougal Dobie}
\author[5]{Natasha Hurley-Walker\orcidlink{0000-0002-5119-4808}}
\author[6,3]{Benito Marcote\orcidlink{000-0001-9814-2354}}
\author[1]{Kaustubh Rajwade\orcidlink{0000-0002-8043-6909}}
\author[6,7]{Antonia Rowlinson\orcidlink{0000-0002-1195-7022}}
\author[8]{Benjamin Stappers\orcidlink{0000-0001-9242-7041}}
\author[9]{Matteo Trudu\orcidlink{0000-0002-1530-0474}}
\author[3,10]{Harish Vedantham}
\author[11,12]{Amanda Weltman\orcidlink{0000-0002-5974-4114}}
\author[]{The Transients Science Working Group}

\affiliation[1]{Astrophysics, Department of Physics, University of Oxford, Denys Wilkinson Building, Keble Road, Oxford OX1 3RH, UK}
\affiliation[2]{Sydney Institute for Astronomy, School of Physics, The University of Sydney, NSW 2006, Australia}
\affiliation[3]{ASTRON, Netherlands Institute for Radio Astronomy, Oude Hoogeveensedijk 4, 7991 PD Dwingeloo, The Netherlands}
\affiliation[4]{Centre for Astrophysics and Supercomputing, Swinburne University of Technology, Hawthorn, VIC 3122, Australia}
\affiliation[5]{International Centre for Radio Astronomy Research, Curtin University, Kent St, Bentley WA 6102, Australia}
\affiliation[6]{Joint Institute for VLBI ERIC, Oude Hoogeveensedijk 4, 7991 PD Dwingeloo, The Netherlands}
\affiliation[7]{Anton Pannekoek Institute for Astronomy, University of Amsterdam, Science Park 904, P.O. Box 94249, 1090GE Amsterdam, The Netherlands}
\affiliation[8]{ Jodrell Bank Centre for Astrophysics, Department of Physics and Astronomy, The University of Manchester, Manchester M13 9PL, UK}
\emailAdd{ben.stappers@manchester.ac.uk}
\affiliation[9]{INAF-Osservatorio Astronomico di Cagliari, via della Scienza 5, 00947 Selargius,Italy}
\affiliation[10]{Kapteyn Astronomical Institute, University of Groningen, PO
Box 72, 97200 AB, Groningen, The Netherlands}
\affiliation[11]{High Energy Physics, Cosmology and Astrophysics Theory (HEPCAT) Group, Department of Mathematics and Applied Mathematics, Universityof Cape Town, Rondebosch, Cape Town, 7700, South Africa}
\affiliation[12]{African Institute for Mathematical Sciences, 6 Melrose Road, Muizenberg, Cape Town, 7945, South Africa}

\abstract{It was already recognised in the 2004 and 2014 SKA Science Case Chapters: "The Exploration of the Unknown" and "The SKA and the Unknown Unknowns", respectively, that the exploration of the unknown is an extremely important concept in realising the potential of any new instrument. This is especially true in an era of limited budgets but expanding analysis algorithms and capabilities. While this Chapter considers transients and environments, the ethos of these previous Chapters should be considered and embraced broadly and can be applied more broadly across SKAO science. We do not repeat in detail the important background these Chapters provide but encourage the reader to review them before reading this Chapter. In the last decade of radio transient astronomy, we have seen the value of these approaches with the expansion of the FRB population, the discovery of long period pulsars and transients and radio emission from white dwarfs to name just a few. With the SKAO, we should push the explored parameter space further and in new directions in all of the sensitivity, field-of-view, time, spectral, spectro-dynamic, polarisation and imaging domains. We should utilise approaches such as machine learning in allowing us to address the complexity that arises from the expanded parameter space and increased data rates and candidate sources. We will briefly summarise some of the impressive progress that the SKA pathfinders and precursors have made in these areas and what these mean for the SKA telescopes. We will also discuss ways in which it might be possible to push the current capabilities.}

\begin{document}
\newcommand{\actaa}{Acta Astron.} 
\newcommand{\araa}{ARA\&A} 
\newcommand{\aar}{A\&ARv} 
\newcommand{\aapr}{A\&ARv} 
\newcommand{\ab}{Astrobiol.} 
\newcommand{\aj}{AJ} 
\newcommand{\apj}{ApJ} 
\newcommand{\apjl}{ApJL} 
\newcommand{\apjs}{ApJSS} 
\newcommand{\ao}{Appl. Opt.} 
\newcommand{\apss}{Astro. \& Space Sci.} 
\newcommand{\aap}{A\&A} 
\newcommand{\aaps}{A\&AS.} 
\newcommand{\baas}{Bull. Am. Astron. Soc.} 
\newcommand{\caa}{Chinese A\&A} 
\newcommand{\cjaa}{Chinese J. A\&A} 
\newcommand{\cqg}{Class. Quantum Gravity} 
\newcommand{\gal}{Galaxies} 
\newcommand{\gca}{Geo. Cosmo. Acta} 
\newcommand{\icarus}{Icarus} 
\newcommand{\jcap}{JCAP} 
\newcommand{\jgr}{J. Geophys. Res.} 
\newcommand{\jgrp}{J. Geophys. Res. Planets} 
\newcommand{\jqsrt}{J. Quant. Spectrosc. Radiat. Transf.} 
\newcommand{\memsai}{Mem. SAIt} 
\newcommand{\mnras}{MNRAS} 
\newcommand{\nat}{Nature} 
\newcommand{\nastro}{Nat. Astron.} 
\newcommand{\ncomms}{Nat. Commun.} 
\newcommand{\nphys}{Nat. Phys.} 
\newcommand{\na}{New Astron.} 
\newcommand{\nar}{New Astron. Rev.} 
\newcommand{\physrep}{Phys. Rep.} 
\newcommand{\pra}{Phys. Rev. A} 
\newcommand{\prb}{Phys. Rev. B} 
\newcommand{\prc}{Phys. Rev. C} 
\newcommand{\prd}{Phys. Rev. D} 
\newcommand{\pre}{Phys. Rev. E} 
\newcommand{\prx}{Phys. Rev. X} 
\newcommand{\prl}{Phys. Rev. Let.} 
\newcommand{\psj}{Planet. Sci. J.} 
\newcommand{\planss}{Planet. Space Sci.} 
\newcommand{\pnas}{Proc. Natl Acad. Sci. USA} 
\newcommand{\procspie}{Proc. SPIE} 
\newcommand{\pasa}{PASA} 
\newcommand{\pasj}{PASJ} 
\newcommand{\pasp}{PASP} 
\newcommand{\rmxaa}{RMXAA} 
\newcommand{\sci}{Science} 
\newcommand{\sciadv}{Sci. Adv.} 
\newcommand{\solphys}{Sol. Phys.} 
\newcommand{\sovast}{Soviet Ast.} 
\newcommand{\ssr}{Space Sci. Rev.} 
\newcommand{\uni}{Universe} 

\maketitle

\section{Introduction}

Serendipity has previously been a fundamental factor that has driven progress in transient astrophysics. For example, an experiment that was designed to explore a certain parameter space results in unexpected discoveries.  However it is important to note that this serendipity very often goes hand in hand with an advancement in technology that results in probing a new parameter space and/or the development of a new data processing capability that enables such unexpected discoveries \citep{1984codi.book.....H}. It is in this sense that we can try to {\it plan} for the unknown unknowns by being as open as possible to facilitating the use of these extensions in capabilities and parameter space, even if they cannot be utilised immediately.

Radio astronomy is rich in such serendipity, with two especially pertinent cases in transient discovery specifically that best exemplify this. In 1967, Dame Jocelyn Bell Burnell was undertaking an experiment to study inter-planetary scintillation using a new telescope and a newly developed chart recorder that could sample data at speeds shorter than 100 milliseconds. That led to the discovery of pulsars, the first manifestation of neutron stars seen in the Galaxy \citep{pulsar}. This has led to an, at the time unimaginable, range of fundamental physics and astrophysics which are described in the Chapters of this volume. Some 4 decades later the revision of an old technique, single pulse searches, was being used to find pulsars in the Magellanic Clouds. When combined with the increased parameter space, in the form of the larger dispersion measures that needed to be searched for these extragalactic pulsars, this resulted in the discovery of Fast Radio Bursts \citep[FRBs,][]{2007Sci...318..777L}. Here too the discovery has implications across a wide range of astrophysics \citep{Curtin01.2026.SKA}. 

A more recent example of success in transient searches which has come from keeping an aspect of parameter space open are the (ultra)-long-period transients (LPTs). The development of the precursor interferometers to the SKA telescopes inspired proposals to undertake searches for rapidly varying sources in the image plane. However the requisite compute capacity meant that these searches were either not undertaken at the time or were limited in scope. Now, in some cases more than 2 decades later, improvements in techniques and computing capacity have enabled these searches and has resulted in the discovery of at least a couple of new types of sources with timescales from minutes to hours \citep[see][and references therein]{Miller-Jones01.2026.SKA} as expanded upon below.

\section{History of this topic with in the SKAO context.}

The science case for the SKA telescopes has evolved considerably over the years \citep{2024skabook,2014aska,CARILLI2004979}. First conceived as a "Hydrogen Array" the case as described in 2004 already expanded well beyond that and even more so in the 2014 case. The 2004 case included a Chapter called the Exploration of the Unknown \citep{2004NewAR..48.1551W} which made the clear point that  "If history is any example the excitement of the SKA will not be in the old questions which are answered, but the new questions that will be raised by the new types of observations it will permit". They also discuss in detail a whole range of search methodologies and observing modes. The transient ones we summarise here as they are all still highly relevant and have proved extremely useful in pathfinder and precursor searches. 

\begin{description}
    \item[Slow Transient Searches:] These are imaging based and can be utilised to find sources which vary on timescales from seconds, minutes and hours (e.g. the newly discovered LPTs), to days for sources like GRB afterglows and emission from X-ray binaries. They also provide the possibility of rapid, repeated, short timescale all-sky surveys.
    \item[Fast Transient Searches:] For these we need to maximise the field of view through appropriate choices of baselines and numbers of coherent beams while maintaining sensitivity. They also require consideration of the effects of dispersion and the associated computational resources (e.g. the expanded population of FRBs).
    \item[Incoherent Summing Modes:] The combination of the antennas/stations incoherently gives the widest field-of-view for that instrument but reduced sensitivity, however, it does allow for use of all available antennas/stations independent of baseline. When combined with the coherent beams these will allow the SKAO to probe both the brighter and fainter end of the luminosity distributions of transients (e.g. both Galactic neutron stars/magnetars and FRBs). 
\end{description}

An important component recognised by \cite{2004NewAR..48.1551W} was the value in ensuring that the developers of the telescopes have vision of the flexibility of the instrument they are building and wherever possible not to preclude a mode or capability. They also noted that the users of the telescope be trained to be familiar enough that they are able to spot anything that might represent a new astrophysical phenomenon. We strongly support this view, but also acknowledge the challenges of the complexity of the SKA telescopes, data processing and data volume and consider in this Chapter the important role of new data analysis approaches but advocate that training and documentation be provided which enables a deeper understanding of the end-to-end data flow. 

In the follow up Chapter in the 2014 Science book \cite{2015aska.confE..65W} noted three SKA Level~0 Science Requirements that had been captured in the document from \cite{2014aska}: the capability to carry out small-scale experimental observations utilising sub-arrays, the capability to rapidly reconfigure the observing mode based on triggers and to enable commensal observations. These are still very important considerations for the SKA telescopes and should continue to be supported. \cite{2015aska.confE..65W} was already concerned that they might be lost when cost pressures came to bear.

\begin{figure}[h]
    \centering
\includegraphics[width=0.8\columnwidth]{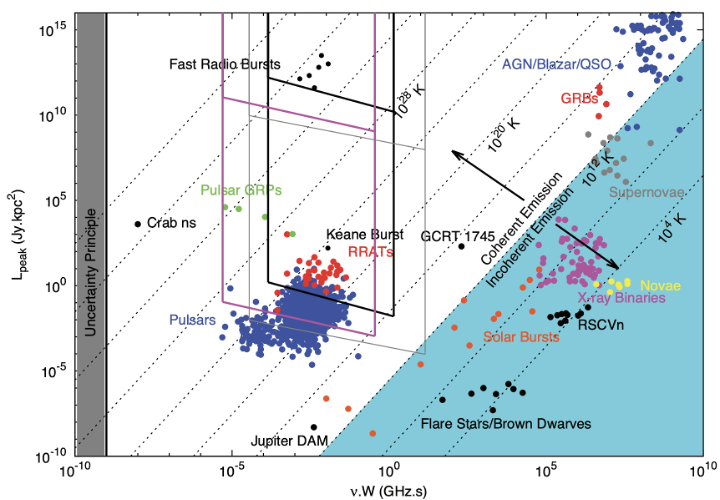}
\includegraphics[width=0.83\columnwidth]{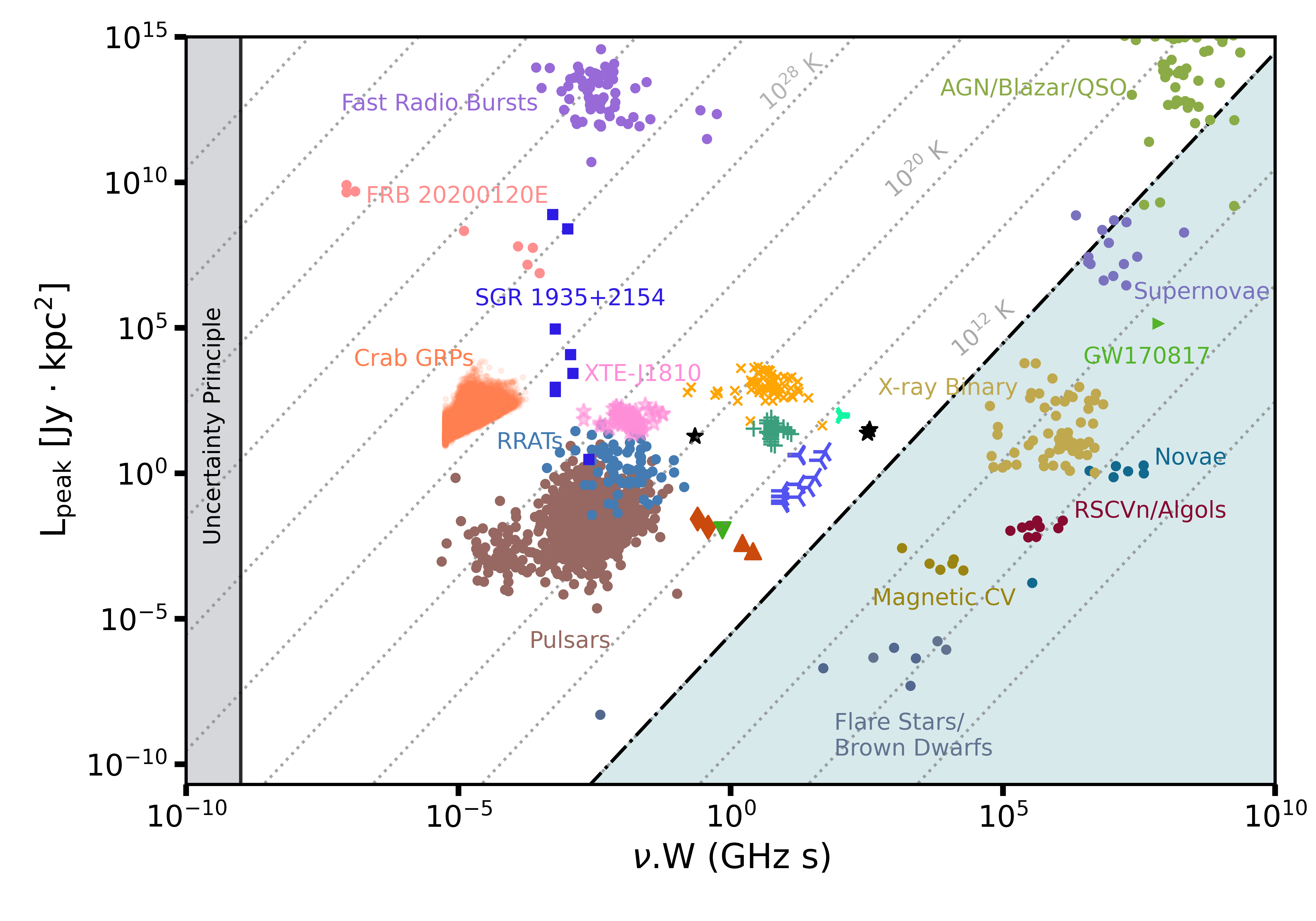}
    \caption{Two representations of the transient phase space diagram in a format based on \cite{2015MNRAS.446.3687P} and \cite{2004NewAR..48.1459C}. The top panel is taken from the 2014 Science book Chapter by \cite{2015aska.confE..55M}. The bottom panel is an updated version which demonstrates the progress made in expanding the parameter space with the SKA pathfinders and precursors.}
    \label{fig:phase_space}
\end{figure}

\section{The motivation for parameter space expansion, recent examples and future extensions.}

\cite{2015aska.confE..65W}, building on the considerations by \cite{1984codi.book.....H}, already did an excellent job of identifying a large range of parameters that should be considered for exploration. Alongside those that are clear from the transient phase space diagram in Figure \ref{fig:phase_space}, i.e. time resolution and luminosity there are some others which are particularly important for transient searches. These are polarised intensity (e.g. LPTs and planet-star interactions), spectral patterns (e.g. FRBs and stars), spectral resolution (again e.g. FRBs), temporal patterns (e.g. stellar flares, X-ray binaries) and observing time which, for example, allows one to detect transients that repeat. \cite{2015aska.confE..65W} also noted the importance of the availability of past observations, and their value has again been borne out with the studies of LPTs, for example. Here we also note that for transient discovery and study maximising the field-of-view and on-sky time are crucial. This has been clearly demonstrated by the SKA precursor and pathfinder discoveries and we will consider below a few of the newer/expanded populations that have been unveiled since the 2014 science book. It is not possible for us to do a full review of this topic, and so refer the reader to the other Chapters in this book on transients \citep{Miller-Jones01.2026.SKA}: Gamma-ray Bursts and Kilonovae from Gravitational Wave Events \citep{Colombo01.2026.SKA}, Interacting Compact Object Binaries \citep{Beri01.2026.SKA}, Tidal Disruption Events and Transients from the Centres of Galaxies \citep{TaoAn03.2026.SKA}, Fast Radio Bursts \citep{Curtin01.2026.SKA,Caleb02.2026.SKA}, Active Galactic Nuclei \citep{Bignall01.2026.SKA}, Supernovae \citep{Chandra01.2026.SKA}, Novae and Symbiotic Stars \citep{Lico01.2026.SKA}, Long Period Transients \citep{Qiu01.2026.SKA} and Stars \citep{Driessen01.2026.SKA} and the recent comprehensive review and outlook on image plane transients by \cite{2025arXiv251110785M} for more details. 

Figure \ref{fig:phase_space} shows two representations of the transient phase space diagram. The top panel was presented in the 2014 science book by \cite{2015aska.confE..55M} which includes the first few FRBs, the first LPTs/GCRTs \citep[GCRT\,J1745$-$3009,][]{2006ApJ...639..348H}. It was mainly used then to highlight the range of parameter space which is still left to explore.  The lower panel shows the excellent progress in expanding the parameter space that has been made using the SKA precursors and pathfinders. There has been a dramatic increase in the number of FRBs, the LPTs are now a population and we have started to fill in the gap between the pulsars and the FRBs with lower luminosity FRBs and very luminous radio pulses from the magnetars. There are also expansions of the populations of X-ray binaries \citep{Beri01.2026.SKA}, radio emitting stars \citep{Driessen01.2026.SKA} and other sources summarised in \cite{Miller-Jones01.2026.SKA} and discussed in the aforementioned chapters in this volume. It is clear though that there is still space that can and should be probed. Noting also from the wide range of possible origins for FRBs that have been proposed \citep{platts_2019_frbmodels} that there are many other types of potential sources to be revealed. 

The higher time-resolution searches that will be possible with SKA-Mid will allow us to push to the left in the transient phase space diagram. High time-resolution searches with SKA-Low will also be interesting, but limited to the nearest by sources due to scattering. The combination of the wider pulse width searches with the image plane searches will push us to the right, and even more so when combined with the higher frequency capabilities of SKA-Mid. The combination of wide field-of-view and sensitivity of both telescopes will allow us to probe an increased volume of space as well as a large area, accessing both the rare bright things and the fainter more common events. There are also likely populations of objects for which we just haven't had the sensitivity to discover them at their given distances and so whole new populations could become apparent. 

The availability of the transient buffer capability for the SKA precursors has already proved extremely valuable \citep[e.g.][]{2019Sci...365..565B,2021ApJ...910..147M,2019Natur.572..352R,2024MNRAS.532.3881R}  and expanding this capability for SKA telescopes for both internal and external triggers will extend this capability. In particular for triggers from other facilities, e.g. high-energy, gravitational wave and optical/IR wide-area monitors \citep[see also the rapid response Chapter,][]{GemmaAnderson01.2026.SKA}, the capability to "look back in time" could probe prompt emission and provide vital important constraints on the astrophysics which is discussed further below and allow the search for unknown unknowns, by providing offline data that can be studied in more detail. 

We now consider three source populations which have been newly uncovered, or greatly expanded with the SKA precursors and pathfinders and also consider the value of opening up the polarisation search dimension. 

\subsection{The long-period radio pulsars}\label{sec:LPPs}
 The recent detection of pulsars with very long rotation periods of \citep{caleb_2022_76spsr,2025ApJ...982L..53W,2025ApJ...990L..49D} previously thought implausible, has expanded the known parameter space of neutron star radio emission dramatically (see \cite{Qiu01.2026.SKA} for details). Their appearance in surveys with existing pathfinder facilities demonstrates how much of the slow-pulsar discovery space remains unexplored. With the vastly improved sensitivity, survey speed, and low-frequency coverage of the SKA telescopes, combined with improved techniques \citep[e.g the FFA;][]{2020MNRAS.497.4654M}, we will be able to probe for even longer-period pulsars and related exotic neutron star populations, almost certainly uncovering new, unexpected classes of radio emitters.

\subsection{The Fast Radio Bursts}

As mentioned above, the SKA pathfinders and precursors enabled giant steps forward for FRB science (see Chapters \citep{Curtin01.2026.SKA,Caleb02.2026.SKA}). 
Of particular interest for the SKA telescopes will be pushing the discovery potential into reaching higher time resolution as well as sensitive image plane searches to extend the variety and sample of FRBs. With the enhanced sensitivity and localisation precision of the SKA telescopes and redshift estimation enabled by the JWST and Rubin observatories. While the sensitivity of SKA-Low should allow us to find a greater population of nearby FRBs which may have multi-wavelength counterparts \citep[see the Chapter on FRB Cosmology for more details,][]{Caleb02.2026.SKA}.

\subsection{The long-period radio transients}\label{sec:LPTs}

In 2002 the discovery of GCRT\,J1745$-$3009 hinted at a new population of radio transients \citep{2006ApJ...639..348H} but as discussed in the LPT Chapter \citep{Qiu01.2026.SKA} it was only with the new, ``fast imaging'' searches that these new ``long-period radio transients'' were established and they may have multiple progenitors. The SKAO will likely detect more both via dedicated and commensal searches (see the commensal searches chapter \citep{AlexAndersson01.2026.SKA}), and be a powerful instrument for follow-up (see the rapid response Chapter \citep{GemmaAnderson01.2026.SKA}).

\subsection{Surveys in polarisation}\label{sec:StokesV}

A further window opened by the SKA precursors is the expansion of untargetted surveys in circular polarisation (e.g. MWA \citep{2018MNRAS.478.2835L}, LOFAR \citep{2021NatAs...5.1233C,2023A&A...670A.124C}, and ASKAP \citep{2021MNRAS.502.5438P,2022MNRAS.516.5972W}). These efforts have resulted in the discovery of new pulsars \citep[e.g.][]{2019ApJ...884...96K,2024ApJ...961..175W,2025arXiv250920611S}, an LPT \citep{2025A&A...699A.341B} and stellar emission (e.g. radio-first discovery of a cold brown dwarf \citep{2020ApJ...903L..33V}), along with linear polarisation from Main-sequence Radio Pulse emitters \citep[MRPs;][]{2025PASA...42..110D}, and tantalising hints of star--exo-planet interactions \citep[SPI;][]{2020NatAs...4..577V,2025A&A...700A.140Z}. See the Chapter on the variability of stars \citep{Driessen01.2026.SKA} for a more detailed discussion. 

Another set of unusual objects that produce circular polarisation have been found using more traditional search methods but more could be found using circular polarisation searches. Examples include GCRT\,J1746$-$2757 \citep{2002AJ....123.1497H}, GCRT\,J1742$-$3001 \citep{2009ApJ...696..280H}, ASKAP\,J173608.2$-$321635 \citep{2021ApJ...920...45W}, VAST J172841.2$-$334548, and VAST J174917.3$-$204841 \citep{2022MNRAS.516.5972W}.

Given the paucity of interfering sources in the circularly polarised sky, transient searches of circularly polarised images are emerging as a convenient way to select for coherent Galactic transients with the SKAO.

\section{Expanding the parameter space}

Radio astronomers now probe an expansive transient parameter space, defined by time, frequency, spatial scale, and all senses of electric field polarisation. This progress is the result of significant advances in instrumentation and techniques, including the development of wideband receivers and high-throughput digital backends, which have pushed the boundaries along each of these dimensions, unveiling new classes of radio transients and deepening our understanding of known phenomena. Unsurprisingly, advances along one axis often enable or sharpen insights along the others: for example, faster time sampling uncovered millisecond bursts that polarisation analysis then linked to extreme, magnetised environments (\cite{michilli18}), while higher spatial resolution secured host-galaxy associations for such events (\cite{chatterjee17}). Below, we discuss each dimension of this multi-dimensional parameter space in turn, highlighting key examples and the interplay between domains.

{\bf Time}. A striking feature of the radio transient zoo is the enormous range of temporal scales, spanning over 17 orders of magnitude. Progress rests on three methodological pillars: {\it high time resolution} to resolve intrinsic structure, {\it improving sensitivity for transients with durations seconds to minutes} to capture longer period sources, and {\it long-term monitoring} to chart long-term variability. High time resolution has exposed rich sub-structure in FRBs \citep{hessels19,nimmo_2022_20Enanoshots,sand_2025_baseband}, with some exhibiting hints of quasi-periodic structure \citep{pm_2023_frbquasi} akin to magnetars \citep{caleb_2022_76spsr, kramer_2024_quasiperiod}, strongly supporting a neutron-star origin. Long-term monitoring has enabled the study of synchrotron-powered transients such as supernovae and GRB afterglows \citep[e.g.][]{weiler_2002,granot_2014} and SN 1993J for example \citep{weiler_2007_sn1993j,martividal_2011_sn1993j2}.  These slow variables—usually found via triggered follow-up—were broadly anticipated. Unexpected, however, and revealed only through searches for wider pulses, were the ultra-long-period coherent transients, for which archival data have been pivotal, as in the case of GPM\,J1839-10 \citep{2023Natur.619..487H}, see the Chapter on longer period transients \citep{Qiu01.2026.SKA} for more details.

{\bf Frequency}.  The spectral behaviour of radio transients is a second, critical axis of their characterisation. Sources can be grouped by whether their emission is intrinsically narrow-band or broadband, and by the richness of their frequency-dependent structure. Progress along this axis has come from wide frequency coverage, which boosts instantaneous sensitivity and reveals unusual spectral windows in both known and newly discovered sources, and from high spectral resolution, which is essential for resolving fine structure and disentangling intrinsic features from propagation effects. 

Wide fractional-band observations and simultaneous multi-band observations indicate that many FRBs are band-limited. This behaviour is, for example, seen in observations with the Ultra-Wideband receivers and by using multi-frequency sub-arraying with MeerKAT \citep{2025MNRAS.540.1685T} and appears more prevalent among repeaters \citep{pleunis_2021_frbmoprh}. Crucially, these insights also motivated sub-band search strategies \citep{kumarsub,trudu_2024_sbs} that explicitly target bursts confined to limited frequency ranges. 

The recent identification of a Type II radio burst with a limited spectral extent in the LOFAR high band from an M dwarf, which is indicative of a coronal mass ejection \citep{2025Natur.647..603C}, also highlights the value of a broad band, but also of sophisticated searches in the dynamic spectra of stars and other targets to find new and interesting emitters.

LPTs have been seen to be brightest in different specific frequency bands \citep[e.g.][]{2022Natur.601..526H,2024ApJ...976L..21H,2025NatAs...9..393L,2025Natur.642..583W}; while it is currently unclear whether this is an intrinsic (emission mechanism) or extrinsic (absorption along the line-of-sight) effect, it motivates untargetted searches across both SKA Low and all Mid bands. 

The wide frequency coverage of the SKA telescopes will be instrumental in extracting the maximum science from incoherent synchrotron emission sources, such as GRBs and accreting compact objects with masses from stellar to super massive \citep[e.g][]{2015MNRAS.446.3687P}. Often the peak emission in these sources evolves with frequency and so a range of frequencies observed (quasi-) simultaneously are crucial to understand the emission processes and viewing geometry. The possibility of sub-arraying using SKA Mid and potentially simultaneous observing with SKA Low, or the use of other SKA precursor and pathfinder telescopes, therefore provides exciting opportunities. For more details on these sources we refer the reader again to other Chapters in this volume, \cite{Beri01.2026.SKA,TaoAn03.2026.SKA,Colombo01.2026.SKA,Driessen01.2026.SKA}.

The unprecedented sensitivity of the SKA at meter-wavelengths is a unique asset because we expect a class of objects (non-relativistic magnetospheres) that have hard spectral cutoffs making them unobservable at higher frequencies. The classic example is the cyclotron emission of Jupiter that has a hard cut off around 40\,MHz. More massive gas-giant planets and planetary mass brown dwarfs are expected to have their cut-offs in the SKA-low band \citep{2010A&A...522A..13R}.
Currently, planetary mass brown dwarfs are identified via their thermal emission but due to the  $T^4$ dependence of thermal luminosity, it is current not feasible to discover the population of (isolated) planetary mass objects below $\approx 300\,$K \citep{2024AJ....167....5L}.
Radio emission being non-thermal in origin need not follow the precipitous luminosity drop with temperature and the sensitivity of the SKA-Low may open up this discovery space. 

{\bf Polarisation}. Retaining the full polarization information yields valuable diagnostics of the magneto-ionic environment surrounding the transient's source and along the line of sight. As noted in Section \ref{sec:StokesV} it also enables discovery of highly polarised emitters that are more readily detected in Q/U/V, which are often more resilient to terrestrial RFI. For instance, polarimetric studies have revealed that many FRBs are highly linearly polarised \citep[often tens of per cent and in some cases approaching $\sim100$\,\%,][]{michilli18}, with occasional but typically modest circular polarisation \citep{kumarcircular}. Getting the polarisation properties of one-off bursts will be key and searches in polarisation space may reveal new FRBs.

Depending on the emission mechanism we are likely to see polarised emission. Gyrosynchrotron emission is expected to result in circular polarisation and for electron cyclotron maser emission we also expect high degrees of circular polarisation, up to 100\%, and these mechanisms are known to be active in stellar systems (see \cite{Driessen01.2026.SKA}
in this volume for more details). 

Further detailed study of the polarisation properties of all transient sources discussed in this volume, especially those that may be one off will be essential. This can be achieved by storing the polarisation data products from the original detection. The analysis  of data in the transient buffers will allow greater fidelity polarisation studies, i.e. with improved calibration, and at higher spectral and temporal resolution. This will be especially important for the unknown unknowns where extracting the maximum data from any detection will allow for more detailed comparison with source models. 

\section{Commensal observing}

The importance of commensal observing for the discovery of transients in general, but for the discovery of ``unknown unknowns" in particular, cannot be overstated. It has been at the core of the discussion around transients \citep[e.g][]{2015aska.confE..51F} and the unknown unknowns \citep[e.g][]{2004NewAR..48.1551W, 2015aska.confE..65W} with the SKA telescopes for decades. By definition, it is impossible to design a search strategy optimised to discover objects whose properties are unknown. Commensal searches offer two key advantages: the ability to access large amounts of observing time (essential for discovering rare events), and the probing of regions of parameter space that would not be chosen for dedicated observations. While commensal searches cannot completely eliminate selection bias, they can significantly {\em reduce} biases across important axes of parameter space such as observing frequency and sky area. As already highlighted unanticipated discoveries made as a result of commensal observing include pulsars, LPTs and FRBs.

A detailed discussion of the techniques and motivation for image plane surveys is discussed in the chapter on commensal imaging in this volume \citep{AlexAndersson01.2026.SKA} and that for the fast transients cases (using both beamforming and dedispersed (where/if needed) imaging searches) can be found in the Chapters \cite{GemmaAnderson01.2026.SKA,Curtin01.2026.SKA} and \cite{Qiu01.2026.SKA} as well as in publications such as \citet{Rajwade2022} and \citet{CRAC0-110ms}.

While commensal observing provides access to ``free" observing time, it is important to recognise that the {\em total} cost of commensal observing is non-zero. Cost categories include computational (to produce and/or analyse the additional data products), storage (scratch space while processing is occurring, or long term archiving of the final commensal data products), calibration overhead (if the commensal observations have different/additional requirements to the primary science goal), and personnel (the time requited to plan, monitor, and exploit the commensal observations). However, with proper planning, these costs can be held low relative to the science return.

Further counterbalancing these costs are the additional non-science benefits of commensal data products. For instance, the high time and frequency resolution used for commensal transient searches can highlight error sources such as narrow-band or low temporal occupancy RFI that corrupts primary science data products at a level that is material yet difficult to diagnose in the time- and frequency-averaged final data. Once identified, they can be masked or (even better) the source identified and eliminated. Instrumental effects (due to calibration, missed packets, or the like) are similarly prominent in high-resolution data and can be identified and the underlying cause corrected, improving the data quality for all science users. A well-rounded commensal program aligned with the reporting and quality assurance procedures leverages these capabilities to maximise the investment in commensal capabilities beyond those purely gained from the commensal science.

\section{Multi-wavelength and Multi-messenger approaches}

Time domain astronomy has entered an era where we no longer focus on single wavelengths, and instead study transients using the entirety of the electromagnetic spectrum, neutrinos and gravitational waves. It has long been the case that many of the transients that are studied in the radio are discovered at other wavelengths \cite[see e.g.][]{2016mks..confE..13F,2025arXiv251110785M} and that will definitely continue into the SKAO era. However, the sensitivity and wide-field-of-view means that one might expect that some of these sources might also be seen initially (e.g. through prompt emission, or afterglows) in the radio. The prompt channels are in particular interesting as they may be indicative of new sources and/or new physics. They in turn would trigger follow up across the multi-wavelength and multi-messenger phase space. The past decade has yielded a variety of ``unknown unknowns'' that demonstrate the need for a holistic approach and we consider some of them here.

The detection of the optical transient AT2018cow led to the establishment of the class of ``Fast Blue Optical Transients'' (FBOTs) which are relativistic transients of as-yet unknown origin. Radio observations from cm to mm wavelengths \citep[e.g][]{2019ApJ...871...73H} were vital in establishing their relativistic nature and will remain key in determining their underlying origin. The Einstein Probe (EP) mission, with its large field of view and rapid localisation capabilities has also led to the establishment of the class of Fast X-ray Transients (FXTs). Whilst consensus is not yet reached, the radio-study of FXTs has constrained their outflow energetics and Lorentz factors \citep[e.g.][]{Shu2025_FXT_mildly_rel}, indicating that some FXTs demonstrate moderately relativistic outflows, indicating that they may originate from the deaths of massive stars similar to gamma-ray bursts. Similarly, revealing the nature of the recently discovered class of Long Period Transients has required comprehensive UV, optical, infrared and X-ray observations to supplement the information gained from their radio emission \citep[see the associated Chapter in this volume;][]{Qiu01.2026.SKA}). In the case of FRBs, beyond the obvious need to locate the host galaxies, detecting a multi-wavelength counterpart to an extragalactic FRB would be a very important component in determining their origin(s). To that end truly simultaneous observations would be needed (see below) for the one-off sources, but follow up or externally triggered observations would be needed for repeating FRBs. 

These four classes of ``unknown unknowns`` demonstrate the symbiotic relationship between different observing wavelengths. This relationship will only escalate in the coming decade with a variety of new facilities coming online that will completely revolutionise the transient parameter space at their retrospective wavelengths. The radio follow up of the afterglow from the neutron star merger GW170817 \citep[e.g][]{2017Sci...358.1579H,2018ApJ...858L..15D} proved crucial for putting constraints on the structure, opening-angle, and viewing geometry. Prompt follow up of more of these GW sources in the radio will be vital for improving our understanding of these systems (see also \cite{Colombo01.2026.SKA,GemmaAnderson01.2026.SKA} in this volume). 

There are a large number of facilities that have just, or will be coming online in the SKAO era, that will be revolutionising the transient phase-space in their particular waveband or messenger type. It isn't possible to discuss them all, but we will briefly consider a few. 

Chapter 8 of the science book for the The Large Synoptic Survey Telescope \citep{2009arXiv0912.0201L}, makes the case for studying the transient and variable Universe. Of particular interest here are the objects that explode, and the variable stars. In both cases radio follow up of as many discoveries as possible will be very valuable, not only to study the known (knowns and unknowns) but also to search for the unknown unknowns. Wide field of view, rapid follow up, sub-arraying and commensal observing will all be useful attributes of the SKA telescopes to match the large area of sky that will be covered with the LSST. 

Synergies between radio and X-ray observatories allow us to probe extremely energetic events. As discussed in the joint white paper on (New)Athena and SKA(O) \citep{2018arXiv180709080C}. Amongst other things they allow one to study the relationship between the energy released and the properties of the accretion (e.g. rate, geometry, optical depth and temperature) with combined radio and X-ray observations. NewAthena should be operating on a timescale commensurate with the SKAO, but earlier (combined) observations with the existing X-ray telescopes will also be vital for progressing this field, as has been possible for example with the ThunderKAT \citep{2016mks..confE..13F} project on the SKA precursor MeerKAT. 

The Nancy Grace Roman Space Telescope will be conducting two transient surveys that will be particularly interesting for the potential for unknown unknowns. The High Latitude Time Domain Survey\footnote{\href{https://science.nasa.gov/mission/roman-space-telescope/high-latitude-time-domain-survey/}{High Latitude Time Domain Survey}} will capture supernovae, TDEs, and merging neutron stars, while the Galactic Bulge Time Domain Survey\footnote{\href{https://science.nasa.gov/mission/roman-space-telescope/galactic-bulge-time-domain-survey/}{Galactic Bulge Time Domain Survey}} will search for exoplanets, brown dwarfs, stellar mass black holes and neutron stars via their microlensing signatures. Both surveys will naturally probe the unknown. Radio follow up of the huge numbers of sources expected to be unveiled by the Nancy Grace Roman Space Telescope will not only reveal unusual manifestations of known sources, but also reveal completely new source classes. 

Co-operation between these facilities and many others we do not have the space to discuss, will yield science results that are greater than the sum of their parts. SKAO capabilities such as automated triggering and transient buffers will allow us to obtain detailed radio information about transient and variables as they occur, and conversely, automated SKAO transient alerts will result in detailed multi-wavelength information by triggering external facilities. Direct coordination in the form of simultaneous observing campaigns will also yield ``unknown unknowns'' \citep[e.g.][]{2019IAUS..339..135A}. Pushing this further, having in place a dedicated co-observing facility such as the MeerKAT and MeerLICHT \citep{2024PASP..136k5003G} combination will allow one to take advantage of the commensal observing with the SKA telescopes and enhance the science output. This will be very useful as some classes of radio transients need contemporaneous (or near contemporaneous) optical band data for secure categorisation \citep[e.g][]{2006ApJ...639..331G}. Moreover this co-observing can also be crucial given the dispersive delay for some sources will mean the radio arrives after the multi-wavelength/multi-messenger signal. The unbiased approach of co-observing provides an opportunity to immediately get a detailed data set on transients and variables "as they occur". This parameter space is relatively unexplored \citep[e.g.][]{10.1093/mnras/stac2143,2023MNRAS.519.4684D,2024MNRAS.531.4836F,2025MNRAS.538.1800H}, and the sensitivity and infrastructure provided by next generation facilities is necessary to probe them comprehensively.

\section{Data processing and machine learning.}

It is recognised that the data volumes that will be generated with the SKA telescopes are so vast that we cannot keep everything. This means that careful consideration should be given to understanding and reviewing what should and will be kept. This requires careful dialogue between the SKAO, the Science Regional Centres and the Science Working Groups. Similarly careful consideration should be given to what searches can be performed in real time and which can be run offline. There must also be a clear pathway for how improvements in any of the relevant algorithms, including machine learning, are implemented, especially in the real time systems. Similarly we must consider how this can be done both efficiently and (cost) effectively. As mentioned above, the nature of the data products that are stored needs to be carefully considered as the value of these data products can be immense, as we already discussed for the LPTs. At SKAO scales, machine learning will increasingly become a core component of the scientific workflow rather than a post-processing convenience. The volume, complexity, and latency requirements of SKAO transient searches mean that many key discovery and data-selection decisions will necessarily rely on automated and ML-enabled systems.

As identified in the previous discussion around this topic in the earlier science books \citep[e.g.][]{2015aska.confE..65W}  the deluge of new data will need more human capacity. Since these earlier SKA science books were written, perhaps the most significant change in the search landscape has been the emergence and adoption of modern machine learning and AI methods. These approaches effectively increase the search bandwidth available to astronomers by automating tasks that previously required substantial human inspection, enabling far larger volumes of data and parameter space to be explored than would otherwise be practical. Already back in 2004, work was undertaken in the pulsar field to group data and start to use heuristics and then in 2014 the first machine learning techniques were introduced \citep[e.g.][]{2014MNRAS.443.1651M,2014ApJ...781..117Z} and see \cite{2016MNRAS.459.1104L} for an overview. These techniques then flowed into the fast transient regime where machine learning implementations have been applied to heuristics or images generated after a candidate has been detected above a set threshold in a trial dedispersed time series \citep[e.g][]{Connor2018,2020MNRAS.497.1661A}. The various techniques that have been applied so far in the image plane searches are discussed in the related Chapter in this volume \citep{AlexAndersson01.2026.SKA}.

The EB-scale data coming from the SKA also necessitates the use of low-latency methods for finding transients. An example of this for multi-purpose interferometers can already be seen in the cases of the CRAFT/CRACO \citep{2010PASA...27..272M,CRAC0-110ms} and MeerTRAP backends \citep[][]{Stappers2016_meertrap} employed on SKA precursors ASKAP and MeerKAT respectively. These backends tap into the telescope datastream, commensally searching for fast transients such as FRBs and bright single pulses from pulsars. This effort requires real-time candidate processing and automatic filtering in order to reduce the data rate to reasonable volumes and also to enable the triggering of the crucial transient buffers. While great progress has been made in the use of ML methods to reduce the candidate volume \citep[e.g.][]{2020MNRAS.497.1661A,Hosenie2021} they are typically still not robust enough to trigger external facilities. To truly explore the unknown unknowns further development work in improving these ML-based inference capabilities is threfore needed to distinguish the sources from, in particular, radio frequency interference which can mimic transients of interest. 

When the SKA telescopes are online, the diluvial rates of even the filtered, interesting transient candidates may prove prohibitive for follow-up with multi-wavelength facilities and the SKA telescopes themselves. At this stage, more comprehensive labelling strategies will need to be adopted. This extends the transient detection problem to one where a label is given to the transient, e.g. an orphan GRB afterglow or an LPT. Such initial concepts have been in place in the radio domain for a while, pre-ML \citep[e.g][]{2017MNRAS.471.3788P,2018MNRAS.479.2481S,2021MNRAS.502..206S} and other schemes are already being developed for optical sky surveys in advance of the Vera Rubin Observatory \citep[][noting here again that capturing the vast literature in this area is beyond the scope of this Chapter]{Stevance2025_VRA}. 
The challenge is building up sufficiently large numbers of candidate events to be able to train the ML models sufficiently. However, it is also important that we remain open to the anomalous (unknown unknown) events as well. 

Radio transients are relatively rare, typically present as at most a few percent of the point sources in an image \citep{Driessen2022_variables, Mooley2016_cnss}, whilst in the time domain they may only be seen in a few to tens of hours of observing, and one of tens of thousands of candidates. This makes anomaly detection techniques - an area of machine learning specifically designed to find rare events that are different to the rest of a distribution - apt tools in the hunt for elusive radio transients. Such approaches may prove as important for discovery as traditional targeted searches, particularly in unexplored regions of parameter space where the expected signal population is not yet known.

Furthermore, unsupervised machine learning techniques by definition do not require a training dataset, making them computationally less intense and suitable for immediate use with SKAO data streams. Initial tests in this area by \cite{Andersson2025_AD} compared the recovery of interesting image plane transients by anomaly detection algorithms to the efforts of citizen scientists, finding that fast anomaly detection algorithms are able to recover the sources that volunteers labeled as transients. These unsupervised algorithms reduced the search volume of candidate transients by an order of magnitude and the false positive rates by a factor of five when comparing to standard statistical cuts on the data. This sort of approach is at the core of what the unknown unknowns is all about. As we expand into new parameter spaces they can ensure that we are in a position to be able to recognise new source types amongst the known sources and the obfuscating radio frequency interference. 

We note also that there is an extensive literature on the use of different approaches, both supervised, semi-supervised, and unsupervised approaches, and we should continue to embrace the developments that are happening widely in these areas. One should consider the options to use simulations to train machine learning models for specific theoretical emission patterns, for example. It is unfortunately not possible to do justice to all the different approaches being trialed in the literature, but it is vital that the community and the SKAO and science regional centres (SRCs) keep abreast of the developments which can dramatically reduce compute requirements.

We should also be aware of significant developments in new methods for processing data that take advantage of machine learning approaches to speed up candidate detection and or selection. This can also be key to finding unknown unknowns as it removes the noise that many dozens of spurious (either noise or RFI) candidates that can mask other new interesting sources before they reach the labelling stage. In the area of fast transients for example consideration has been given to searching directly in the so-called DM-t (dispersion measure and time) plane which takes advantage of more information than a standard thresholding technique does \citep[e.g][]{2025ApJS..276...20Z,2025arXiv251119014B}. It can also reduce significantly the number of candidates that need to be clustered and sifted. There are also approaches that use machine learning to more efficiently sift through candidates to avoid the more costly search steps \citep[e.g][]{2025MNRAS.tmp.1841F}.

Expanding the parameter space over which transients are searched for not only greatly expands the data rates but also increases the number of trials that are being used to probe this increased parameter space. It is therefore important that the analysis techniques are able to correctly deal with the statistical properties of these data sets in order to ascertain the validity of any new discoveries. This is going to be most important when we are considering one-off events and/or when associating events across both the photonic and non-photonic regimes. It is also going to be important that we consider carefully the results of our application of our ML approaches and how we interpret the results of those, especially at the limits of sensitivity. 

It is also going to be important that we don't stand still when considering our analysis of the transient data and take advantages of the rapid developments in both algorithms and compute infrastructure. Carefully storing and curating the data products that can be archived from the SKA telescopes will be crucial in enabling improved ML models to be developed and tested. Building an infrastructure the allows the easy and traceable application (to extract new science), testing (to potentially have them be part of the real-time search pipeline) and comparison with previous results will greatly facilitate this. While most important for models which used (semi-)supervised learning approaches, they will also be vital for testing unsupervised and anomaly detection based models. 

\section{Conclusions}

In this Chapter we have re-iterated the importance of the search for unknown unknowns with the SKA telescopes. This concept has been part of the SKAO ethos since the expansion of the SKAO science case in 2004 and was ultimately adopted into the Level~0 science goals in 2014. These captured the ideas of enabling small-scale experimental observations using portions of the telescopes, the ability to rapidly reconfigure based on triggers and to enable commensal observations. As discussed here these are still essential elements of the SKAO for maximising discovery capabilities. Building on the detailed case put forward by \cite{2004NewAR..48.1551W} in "The Exploration of the Unknown", Cordes and the International SKA Science Working Group put out Memo 85 entitled "Discovery and Understanding with the SKA" in 2006 which captured in great detail how to go about achieving these dual goals with the SKA. In the 2014 science case \cite{2015aska.confE..65W} returned to the idea of discovering new classes of astrophysical objects with his "The SKA and the Unknown Unknowns". These documents, and others, are essential reading to review how to best maximise the discovery science possible with the SKA telescopes. We have provided evidence of the value of this approach when applied to the SKA precursors and pathfinders. We have discussed how expanding even further into the corners of parameter space that will be possible with the SKA telescopes will allow us to further pursue the unknown unknowns. We have also considered how it is vital that we embrace the use of tools such as machine learning to not only expand the discovery space but to also pursue anomalies in the data that may reveal new wonders. In this respect, advances in machine learning and automated analysis should be viewed not simply as operational tools, but as a key component of the SKAO's ability to explore previously inaccessible regions of observational parameter space and to recognise genuinely unexpected phenomena. Borrowing from Memo 85 we reiterate that the SKA telescopes and associated analysis software must come together to be a discovery instrument, both for known and unknown classes of astrophysical sources.

\section*{Acknowledgements}

AR acknowledges funding from the European Research Council (ERC) under the European Union’s Horizon research and innovation programme (‘QuickBlitz’; grant agreement number 101170284), funding from the NWO Aspasia grant (number 015.016.033), and support through the project CORTEX (project number NWA.1160.18.316) of the research programme NWA-ORC, which is (partly) financed by the Dutch Research Council (NWO). Views and opinions expressed are however those of the author(s) only and do not necessarily reflect those of the European Union or the European Research Council. Neither the European Union nor the granting authority can be held responsible for them. Part of the research activities described in this paper were carried out with the contribution of the Next Generation EU funds within the National Recovery and Resilience Plan (PNRR), Mission 4 - Education and Research, Component 2 - From Research to Business (M4C2), Investment Line 3.1 - Strengthening and creation of Research Infrastructures, Project IR0000026 – Next Generation Croce del Nord.

\bibliographystyle{abbrvnat-maxbibnames4}
\bibliography{chapter} 

@incollection{Miller-Jones01.2026.SKA, author = {James Miller-Jones and  Kaustubh Rajwade and Patrick A. Woudt and Jason W.T. Hessels},title = {},year = {2026},publisher = {},note = {arXiv search: Report number AASKAII/Miller-Jones01},booktitle = {Advancing Astrophysics with the SKA -- II (AASKAII)}}

@incollection{Driessen01.2026.SKA, author = {Laura N. Driessen and author2 and author3 and author4 and author5},title = {},year = {2026},publisher = {},note = {arXiv search: Report number AASKAII/Driessen01},booktitle = {Advancing Astrophysics with the SKA -- II (AASKAII)}}

@incollection{GemmaAnderson01.2026.SKA, author = {Gemma E. Anderson and author2 and author3 and author4 and author5},title = {},year = {2026},publisher = {},note = {arXiv search: Report number AASKAII/GemmaAnderson01},booktitle = {Advancing Astrophysics with the SKA -- II (AASKAII)}}

@incollection{Qiu01.2026.SKA, author = {Hao Qiu and author2 and author3 and author4 and author5},title = {},year = {2026},publisher = {},note = {arXiv search: Report number AASKAII/Qiu01},booktitle = {Advancing Astrophysics with the SKA -- II (AASKAII)}}

@incollection{Lico01.2026.SKA, author = {Rocco Lico and author2 and author3 and author4 and author5},title = {},year = {2026},publisher = {},note = {arXiv search: Report number AASKAII/Lico01},booktitle = {Advancing Astrophysics with the SKA -- II (AASKAII)}}

@incollection{TaoAn03.2026.SKA, author = {Tao An and author2 and author3 and author4 and author5},title = {},year = {2026},publisher = {},note = {arXiv search: Report number AASKAII/TaoAn03},booktitle = {Advancing Astrophysics with the SKA -- II (AASKAII)}}

@incollection{Beri01.2026.SKA, author = {Aru Beri and author2 and author3 and author4 and author5},title = {},year = {2026},publisher = {},note = {arXiv search: Report number AASKAII/Beri01},booktitle = {Advancing Astrophysics with the SKA -- II (AASKAII)}}

@incollection{Caleb02.2026.SKA, author = {Manisha Caleb and author2 and author3 and author4 and author5},title = {},year = {2026},publisher = {},note = {arXiv search: Report number AASKAII/Caleb02},booktitle = {Advancing Astrophysics with the SKA -- II (AASKAII)}}

@incollection{Bignall01.2026.SKA, author = {Hayley Bignall and author2 and author3 and author4 and author5},title = {},year = {2026},publisher = {},note = {arXiv search: Report number AASKAII/Bignall01},booktitle = {Advancing Astrophysics with the SKA -- II (AASKAII)}}

@incollection{AlexAndersson01.2026.SKA, author = {Alex Andersson and author2 and author3 and author4 and author5},title = {},year = {2026},publisher = {},note = {arXiv search: Report number AASKAII/AlexAndersson01},booktitle = {Advancing Astrophysics with the SKA -- II (AASKAII)}}

@incollection{Colombo01.2026.SKA, author = {Alberto Colombo and author2 and author3 and author4 and author5},title = {},year = {2026},publisher = {},note = {arXiv search: Report number AASKAII/Colombo01},booktitle = {Advancing Astrophysics with the SKA -- II (AASKAII)}}

@incollection{Chandra01.2026.SKA, author = {Poonam Chandra and author2 and author3 and author4 and author5},title = {},year = {2026},publisher = {},note = {arXiv search: Report number AASKAII/Chandra01},booktitle = {Advancing Astrophysics with the SKA -- II (AASKAII)}}

@incollection{Curtin01.2026.SKA, author = {Alice P. Curtin and author2 and author3 and author4 and author5},title = {},year = {2026},publisher = {},note = {arXiv search: Report number AASKAII/Curtin01},booktitle = {Advancing Astrophysics with the SKA -- II (AASKAII)}}

@ARTICLE{2020MNRAS.497.4654M,
       author = {{Morello}, V. and {Barr}, E.~D. and {Stappers}, B.~W. and {Keane}, E.~F. and {Lyne}, A.~G.},
        title = "{Optimal periodicity searching: revisiting the fast folding algorithm for large-scale pulsar surveys}",
      journal = {\mnras},
         year = 2020,
        month = oct,
       volume = {497},
       number = {4},
        pages = {4654-4671},
          doi = {10.1093/mnras/staa2291},
archivePrefix = {arXiv},
       eprint = {2004.03701},
 primaryClass = {astro-ph.IM},
       adsurl = {https://ui.adsabs.harvard.edu/abs/2020MNRAS.497.4654M}
}

@ARTICLE{2016MNRAS.459.1104L,
       author = {{Lyon}, R.~J. and {Stappers}, B.~W. and {Cooper}, S. and {Brooke}, J.~M. and {Knowles}, J.~D.},
        title = "{Fifty Years of Pulsar Candidate Selection: From simple filters to a new principled real-time classification approach}",
      journal = {\mnras},
         year = 2016,
        month = jun,
       volume = {459},
        pages = {1104},
          doi = {10.1093/mnras/stw656},
archivePrefix = {arXiv},
       eprint = {1603.05166},
 primaryClass = {astro-ph.IM},
       adsurl = {https://ui.adsabs.harvard.edu/abs/2016MNRAS.459.1104L}
}

@ARTICLE{2014ApJ...781..117Z,
       author = {{Zhu}, W.~W. and {Berndsen}, A. and {Madsen}, E.~C. and {Tan}, M. and {Stairs}, I.~H. and {Brazier}, A. and {Lazarus}, P. and {Lynch}, R. and {Scholz}, P. and {Stovall}, K. and {Ransom}, S.~M. and {Banaszak}, S. and {Biwer}, C.~M. and {Cohen}, S. and {Dartez}, L.~P. and {Flanigan}, J. and {Lunsford}, G. and {Martinez}, J.~G. and {Mata}, A. and {Rohr}, M. and {Walker}, A. and {Allen}, B. and {Bhat}, N.~D.~R. and {Bogdanov}, S. and {Camilo}, F. and {Chatterjee}, S. and {Cordes}, J.~M. and {Crawford}, F. and {Deneva}, J.~S. and {Desvignes}, G. and {Ferdman}, R.~D. and {Freire}, P.~C.~C. and {Hessels}, J.~W.~T. and {Jenet}, F.~A. and {Kaplan}, D.~L. and {Kaspi}, V.~M. and {Knispel}, B. and {Lee}, K.~J. and {van Leeuwen}, J. and {Lyne}, A.~G. and {McLaughlin}, M.~A. and {Siemens}, X. and {Spitler}, L.~G. and {Venkataraman}, A.},
        title = "{Searching for Pulsars Using Image Pattern Recognition}",
      journal = {\apj},
         year = 2014,
        month = feb,
       volume = {781},
       number = {2},
          eid = {117},
        pages = {117},
          doi = {10.1088/0004-637X/781/2/117},
archivePrefix = {arXiv},
       eprint = {1309.0776},
 primaryClass = {astro-ph.IM},
       adsurl = {https://ui.adsabs.harvard.edu/abs/2014ApJ...781..117Z}
}

@ARTICLE{2014MNRAS.443.1651M,
       author = {{Morello}, V. and {Barr}, E.~D. and {Bailes}, M. and {Flynn}, C.~M. and {Keane}, E.~F. and {van Straten}, W.},
        title = "{SPINN: a straightforward machine learning solution to the pulsar candidate selection problem}",
      journal = {\mnras},
         year = 2014,
        month = sep,
       volume = {443},
       number = {2},
        pages = {1651-1662},
          doi = {10.1093/mnras/stu1188},
archivePrefix = {arXiv},
       eprint = {1406.3627},
 primaryClass = {astro-ph.IM},
       adsurl = {https://ui.adsabs.harvard.edu/abs/2014MNRAS.443.1651M}
}

@article{10.1093/mnras/stac2143,
    author = {Rigney, Jeremy and Ramsay, Gavin and Carley, Eoin P and Doyle, J Gerry and Gallagher, Peter T and Wang, Yuanming and Pritchard, Joshua and Murphy, Tara and Lenc, Emil and Kaplan, David L},
    title = {Searching for stellar flares from low-mass stars using ASKAP and TESS},
    journal = {Monthly Notices of the Royal Astronomical Society},
    volume = {516},
    number = {1},
    pages = {540-549},
    year = {2022},
    month = {08},
    issn = {0035-8711},
    doi = {10.1093/mnras/stac2143},
    url = {https://doi.org/10.1093/mnras/stac2143},
    eprint = {https://academic.oup.com/mnras/article-pdf/516/1/540/45583130/stac2143.pdf},
}

@phdthesis{Hosenie2021,
    author = {{Hosenie}, Z.},
    title = {Feature Detection and Classification in streaming and non-streaming astronomical datasets},
    school = {University of Manchester},
    year = {2021}
}

@ARTICLE{2025arXiv251110785M,
       author = {{Murphy}, Tara and {Kaplan}, David L.},
        title = "{The Dawes Review 13: A New Look at The Dynamic Radio Sky}",
      journal = {arXiv e-prints},
         year = 2025,
        month = nov,
          eid = {arXiv:2511.10785},
        pages = {arXiv:2511.10785},
          doi = {10.48550/arXiv.2511.10785},
archivePrefix = {arXiv},
       eprint = {2511.10785},
 primaryClass = {astro-ph.SR},
       adsurl = {https://ui.adsabs.harvard.edu/abs/2025arXiv251110785M}
}

@ARTICLE{2025MNRAS.tmp.1841F,
       author = {{Fu}, Qiuyang and {Xue}, Mengyao and {Zhu}, Weiwei and {Bhat}, N.~D.~R. and {Wu}, Kaichao and {Zhang}, Zihan and {Meyers}, B.~W. and {Tan}, Chia Min and {Yue}, Youling and {Niu}, Jiarui and {Meng}, Lingqi and {Wu}, Ziwei and {Fang}, Ziyao and {Zhou}, Yukai and {Jin}, Jiawei},
        title = "{Improving pulsar search efficiency in next-generation pulsar surveys with artificial intelligence}",
      journal = {\mnras},
         year = 2025,
        month = nov,
          doi = {10.1093/mnras/staf1952},
archivePrefix = {arXiv},
       eprint = {2511.06711},
 primaryClass = {astro-ph.IM},
       adsurl = {https://ui.adsabs.harvard.edu/abs/2025MNRAS.tmp.1841F}
}

@ARTICLE{2025ApJS..276...20Z,
       author = {{Zhang}, Yong-Kun and {Li}, Di and {Feng}, Yi and {Tsai}, Chao-Wei and {Wang}, Pei and {Niu}, Chen-Hui and {Chen}, Hua-Xi and {Zhu}, Yu-Hao},
        title = "{DRAFTS: A Deep-learning-based Radio Fast Transient Search Pipeline}",
      journal = {\apjs},
         year = 2025,
        month = jan,
       volume = {276},
       number = {1},
          eid = {20},
        pages = {20},
          doi = {10.3847/1538-4365/ad8f31},
archivePrefix = {arXiv},
       eprint = {2410.03200},
 primaryClass = {astro-ph.IM},
       adsurl = {https://ui.adsabs.harvard.edu/abs/2025ApJS..276...20Z}
}

@ARTICLE{2025arXiv251119014B,
       author = {{Belmonte Diaz}, Sergio and {Breton}, Rene P. and {Hosenie}, Zafiirah and {Stappers}, Ben W.},
        title = "{Enhancing Fast Radio Transient Detection with Mask R-CNN Image Segmentation}",
      journal = {arXiv e-prints},
         year = 2025,
        month = nov,
          eid = {arXiv:2511.19014},
        pages = {arXiv:2511.19014},
          doi = {10.48550/arXiv.2511.19014},
archivePrefix = {arXiv},
       eprint = {2511.19014},
 primaryClass = {astro-ph.IM},
       adsurl = {https://ui.adsabs.harvard.edu/abs/2025arXiv251119014B}
}

@ARTICLE{2021MNRAS.502..206S,
       author = {{Sooknunan}, K. and {Lochner}, M. and {Bassett}, Bruce A. and {Peiris}, H.~V. and {Fender}, R. and {Stewart}, A.~J. and {Pietka}, M. and {Woudt}, P.~A. and {McEwen}, J.~D. and {Lahav}, O.},
        title = "{Classification of multiwavelength transients with machine learning}",
      journal = {\mnras},
         year = 2021,
        month = mar,
       volume = {502},
       number = {1},
        pages = {206-224},
          doi = {10.1093/mnras/staa3873},
archivePrefix = {arXiv},
       eprint = {1811.08446},
 primaryClass = {astro-ph.IM},
       adsurl = {https://ui.adsabs.harvard.edu/abs/2021MNRAS.502..206S}
}

@ARTICLE{2017MNRAS.471.3788P,
       author = {{Pietka}, M. and {Staley}, T.~D. and {Pretorius}, M.~L. and {Fender}, R.~P.},
        title = "{On the use of variability time-scales as an early classifier of radio transients and variables}",
      journal = {\mnras},
         year = 2017,
        month = nov,
       volume = {471},
       number = {4},
        pages = {3788-3805},
          doi = {10.1093/mnras/stx1744},
archivePrefix = {arXiv},
       eprint = {1707.04265},
 primaryClass = {astro-ph.HE},
       adsurl = {https://ui.adsabs.harvard.edu/abs/2017MNRAS.471.3788P}
}

@ARTICLE{2018MNRAS.479.2481S,
       author = {{Stewart}, A.~J. and {Mu{\~n}oz-Darias}, T. and {Fender}, R.~P. and {Pietka}, M.},
        title = "{On the optical counterparts of radio transients and variables}",
      journal = {\mnras},
         year = 2018,
        month = sep,
       volume = {479},
       number = {2},
        pages = {2481-2504},
          doi = {10.1093/mnras/sty1671},
archivePrefix = {arXiv},
       eprint = {1806.09815},
 primaryClass = {astro-ph.HE},
       adsurl = {https://ui.adsabs.harvard.edu/abs/2018MNRAS.479.2481S}
}

@ARTICLE{2023MNRAS.519.4684D,
       author = {{Dobie}, D. and {Pritchard}, J. and {Wang}, Y. and {Graham}, L.~W. and {Freeburn}, J. and {Qiu}, H. and {White}, T.~R. and {O'Brien}, A. and {Lenc}, E. and {Leung}, J.~K. and {Lynch}, C. and {Murphy}, Tara and {Stewart}, A.~J. and {Wang}, Z. and {Zic}, A. and {Abbott}, T.~M.~C. and {Cai}, C. and {Cooke}, J. and {Dobiecki}, M. and {Goode}, S. and {Jia}, S. and {Li}, C. and {M{\"o}ller}, A. and {Webb}, S. and {Zhang}, J. and {Zhang}, S.~N.},
        title = "{Radio transients and variables in the tenth Deeper, Wider, Faster observing run}",
      journal = {\mnras},
         year = 2023,
        month = mar,
       volume = {519},
       number = {3},
        pages = {4684-4698},
          doi = {10.1093/mnras/stac3731},
archivePrefix = {arXiv},
       eprint = {2211.07049},
 primaryClass = {astro-ph.HE},
       adsurl = {https://ui.adsabs.harvard.edu/abs/2023MNRAS.519.4684D}
}

@ARTICLE{2024MNRAS.531.4836F,
       author = {{Freeburn}, James and {Cooke}, Jeff and {M{\"o}ller}, Anais and {Dobie}, Dougal and {Zhang}, Jielai and {Salafia}, Om Sharan and {Siellez}, Karelle and {Auchettl}, Katie and {Goode}, Simon and {Abbott}, Timothy M.~C. and {Andreoni}, Igor and {Allen}, Rebecca and {Van Bemmel}, Natasha and {Webb}, Sara},
        title = "{A fast-cadenced search for gamma-ray burst orphan afterglows with the Deeper, Wider, Faster programme}",
      journal = {\mnras},
         year = 2024,
        month = jul,
       volume = {531},
       number = {4},
        pages = {4836-4851},
          doi = {10.1093/mnras/stae1489},
archivePrefix = {arXiv},
       eprint = {2405.11949},
 primaryClass = {astro-ph.HE},
       adsurl = {https://ui.adsabs.harvard.edu/abs/2024MNRAS.531.4836F}
}

@ARTICLE{2024PASP..136k5003G,
       author = {{Groot}, P.~J. and {Bloemen}, S. and {Vreeswijk}, P.~M. and {van Roestel}, J.~C.~J. and {Jonker}, P.~G. and {Nelemans}, G. and {Klein-Wolt}, M. and {Lepoole}, R. and {Pieterse}, D.~L.~A. and {Rodenhuis}, M. and {Boland}, W. and {Haverkorn}, M. and {Aerts}, C. and {Bakker}, R. and {Balster}, H. and {Bekema}, M. and {Dijkstra}, E. and {Dolron}, P. and {Elswijk}, E. and {van Elteren}, A. and {Engels}, A. and {Fokker}, M. and {de Haan}, M. and {Hahn}, F. and {ter Horst}, R. and {Lesman}, D. and {Kragt}, J. and {Morren}, J. and {Nillissen}, H. and {Pessemier}, W. and {Raskin}, G. and {de Rijke}, A. and {Scheers}, L.~H.~A. and {Schuil}, M. and {Timmer}, S.~T. and {Antunes Amaral}, L. and {Arancibia-Rojas}, E. and {Arcavi}, I. and {Blagorodnova}, N. and {Biswas}, S. and {Breton}, R.~P. and {Dawson}, H. and {Dayal}, P. and {De Wet}, S. and {Duffy}, C. and {Faris}, S. and {Fausnaugh}, M. and {Gal-Yam}, A. and {Geier}, S. and {Horesh}, A. and {Johnston}, C. and {Katusiime}, G. and {Kelley}, C. and {Kosakowski}, A. and {Kupfer}, T. and {Leloudas}, G. and {Levan}, A. and {Modiano}, D. and {Mogawana}, O. and {Munday}, J. and {Paice}, J. and {Patat}, F. and {Pelisoli}, I. and {Ramsay}, G. and {Ranaivomanana}, P.~T. and {Ruiz-Carmona}, R. and {Schaffenroth}, V. and {Scaringi}, S. and {Stoppa}, F. and {Street}, R. and {Tranin}, H. and {Uzundag}, M. and {Valenti}, S. and {Veresvarska}, M. and {Vuc̆kovi{\'c}}, M. and {Wichern}, H.~C.~I. and {Wijers}, R.~A.~M.~J. and {Wijnands}, R.~A.~D. and {Zimmerman}, E.},
        title = "{The BlackGEM Telescope Array. I. Overview}",
      journal = {\pasp},
         year = 2024,
        month = nov,
       volume = {136},
       number = {11},
          eid = {115003},
        pages = {115003},
          doi = {10.1088/1538-3873/ad8b6a},
archivePrefix = {arXiv},
       eprint = {2405.18923},
 primaryClass = {astro-ph.IM},
       adsurl = {https://ui.adsabs.harvard.edu/abs/2024PASP..136k5003G}
}

@ARTICLE{2025MNRAS.538.1800H,
       author = {{Hanmer}, K.~Y. and {Pastor-Marazuela}, I. and {Brink}, J. and {Malesani}, D. and {Stappers}, B.~W. and {Groot}, P.~J. and {Cooper}, A.~J. and {Tejos}, N. and {Buckley}, D.~A.~H. and {Barr}, E.~D. and {Bezuidenhout}, M.~C. and {Bloemen}, S. and {Caleb}, M. and {Driessen}, L.~N. and {Fender}, R. and {Jankowski}, F. and {Kramer}, M. and {Pieterse}, D.~L.~A. and {Rajwade}, K.~M. and {Tian}, J. and {Vreeswijk}, P.~M. and {Wijnands}, R. and {Woudt}, P.~A.},
        title = "{Contemporaneous optical-radio observations of a fast radio burst in a close galaxy pair}",
      journal = {\mnras},
         year = 2025,
        month = apr,
       volume = {538},
       number = {3},
        pages = {1800-1815},
          doi = {10.1093/mnras/staf289},
archivePrefix = {arXiv},
       eprint = {2502.10153},
 primaryClass = {astro-ph.HE},
       adsurl = {https://ui.adsabs.harvard.edu/abs/2025MNRAS.538.1800H}
}

@INPROCEEDINGS{2019IAUS..339..135A,
       author = {{Andreoni}, I. and {Cooke}, J.},
        title = "{The Deeper Wider Faster Programme: Chasing the Fastest Bursts in the Universe}",
    booktitle = {Southern Horizons in Time-Domain Astronomy},
         year = 2019,
       editor = {{Griffin}, R. Elizabeth},
       series = {IAU Symposium},
       volume = {339},
        month = aug,
        pages = {135-138},
          doi = {10.1017/S1743921318002399},
archivePrefix = {arXiv},
       eprint = {1802.01100},
 primaryClass = {astro-ph.IM},
       adsurl = {https://ui.adsabs.harvard.edu/abs/2019IAUS..339..135A}
}

@ARTICLE{2025MNRAS.540.1685T,
       author = {{Tian}, J. and {Pastor-Marazuela}, I. and {Rajwade}, K.~M. and {Stappers}, B.~W. and {Shaji}, K. and {Hanmer}, K.~Y. and {Caleb}, M. and {Bezuidenhout}, M.~C. and {Jankowski}, F. and {Breton}, R. and {Barr}, E.~D. and {Kramer}, M. and {Groot}, P.~J. and {Bloemen}, S. and {Vreeswijk}, P. and {Pieterse}, D. and {Woudt}, P.~A. and {Fender}, R.~P. and {Wijnands}, R.~A.~D. and {Buckley}, D.~A.~H.},
        title = "{MeerKAT discovery of a hyperactive repeating fast radio burst source}",
      journal = {\mnras},
         year = 2025,
        month = jun,
       volume = {540},
       number = {2},
        pages = {1685-1700},
          doi = {10.1093/mnras/staf793},
archivePrefix = {arXiv},
       eprint = {2505.08372},
 primaryClass = {astro-ph.HE},
       adsurl = {https://ui.adsabs.harvard.edu/abs/2025MNRAS.540.1685T}
}

@ARTICLE{2025ApJ...990L..49D,
       author = {{Dong}, Fengqiu Adam and {Clarke}, Tracy E. and {Curtin}, Alice and {Kumar}, Ajay and {Mckinven}, Ryan and {Shin}, Kaitlyn and {Stairs}, Ingrid and {Brar}, Charanjot and {Burdge}, Kevin and {Chatterjee}, Shami and {Cook}, Amanda M. and {Fonseca}, Emmanuel and {Gaensler}, B.~M. and {Hessels}, Jason W. and {Kaspi}, Victoria M. and {Lazda}, Mattias and {Main}, Robert and {Masui}, Kiyoshi W. and {McKee}, James W. and {Meyers}, Bradley W. and {Pearlman}, Aaron B. and {Ransom}, Scott M. and {Scholz}, Paul and {Smith}, Kendrick M. and {Tan}, Chia Min},
        title = "{CHIME/Fast Radio Burst/Pulsar Discovery of a Nearby Long-period Radio Transient with a Timing Glitch}",
      journal = {\apjl},
         year = 2025,
        month = sep,
       volume = {990},
       number = {2},
          eid = {L49},
        pages = {L49},
          doi = {10.3847/2041-8213/adfa8e},
archivePrefix = {arXiv},
       eprint = {2407.07480},
 primaryClass = {astro-ph.HE},
       adsurl = {https://ui.adsabs.harvard.edu/abs/2025ApJ...990L..49D}
}

@ARTICLE{2025ApJ...982L..53W,
       author = {{Wang}, Yuanming and {Uttarkar}, Pavan A. and {Shannon}, Ryan M. and {Lee}, Yu Wing Joshua and {Dobie}, Dougal and {Wang}, Ziteng and {Bannister}, Keith W. and {Caleb}, Manisha and {Deller}, Adam T. and {Glowacki}, Marcin and {Jahns-Schindler}, Joscha N. and {Murphy}, Tara and {Anna-Thomas}, Reshma and {Bhat}, N.~D.~R. and {Deng}, Xinping and {Gupta}, Vivek and {Jaini}, Akhil and {James}, Clancy W. and {Tuthill}, John},
        title = "{The Discovery of a 41 s Radio Pulsar PSR J0311+1402 with ASKAP}",
      journal = {\apjl},
         year = 2025,
        month = apr,
       volume = {982},
       number = {2},
          eid = {L53},
        pages = {L53},
          doi = {10.3847/2041-8213/adbe61},
archivePrefix = {arXiv},
       eprint = {2503.07936},
 primaryClass = {astro-ph.HE},
       adsurl = {https://ui.adsabs.harvard.edu/abs/2025ApJ...982L..53W}
}

@ARTICLE{2019Natur.572..352R,
       author = {{Ravi}, V. and {Catha}, M. and {D'Addario}, L. and {Djorgovski}, S.~G. and {Hallinan}, G. and {Hobbs}, R. and {Kocz}, J. and {Kulkarni}, S.~R. and {Shi}, J. and {Vedantham}, H.~K. and {Weinreb}, S. and {Woody}, D.~P.},
        title = "{A fast radio burst localized to a massive galaxy}",
      journal = {\nat},
         year = 2019,
        month = aug,
       volume = {572},
       number = {7769},
        pages = {352-354},
          doi = {10.1038/s41586-019-1389-7},
archivePrefix = {arXiv},
       eprint = {1907.01542},
 primaryClass = {astro-ph.HE},
       adsurl = {https://ui.adsabs.harvard.edu/abs/2019Natur.572..352R}
}

@ARTICLE{2021ApJ...910..147M,
       author = {{Michilli}, D. and {Masui}, K.~W. and {Mckinven}, R. and {Cubranic}, D. and {Bruneault}, M. and {Brar}, C. and {Patel}, C. and {Boyle}, P.~J. and {Stairs}, I.~H. and {Renard}, A. and {Bandura}, K. and {Berger}, S. and {Breitman}, D. and {Cassanelli}, T. and {Dobbs}, M. and {Kaspi}, V.~M. and {Leung}, C. and {Mena-Parra}, J. and {Pleunis}, Z. and {Russell}, L. and {Scholz}, P. and {Siegel}, S.~R. and {Tendulkar}, S.~P. and {Vanderlinde}, K.},
        title = "{An Analysis Pipeline for CHIME/FRB Full-array Baseband Data}",
      journal = {\apj},
         year = 2021,
        month = apr,
       volume = {910},
       number = {2},
          eid = {147},
        pages = {147},
          doi = {10.3847/1538-4357/abe626},
archivePrefix = {arXiv},
       eprint = {2010.06748},
 primaryClass = {astro-ph.HE},
       adsurl = {https://ui.adsabs.harvard.edu/abs/2021ApJ...910..147M}
}

@ARTICLE{2019Sci...365..565B,
       author = {{Bannister}, K.~W. and {Deller}, A.~T. and {Phillips}, C. and {Macquart}, J.-P. and {Prochaska}, J.~X. and {Tejos}, N. and {Ryder}, S.~D. and {Sadler}, E.~M. and {Shannon}, R.~M. and {Simha}, S. and {Day}, C.~K. and {McQuinn}, M. and {North-Hickey}, F.~O. and {Bhandari}, S. and {Arcus}, W.~R. and {Bennert}, V.~N. and {Burchett}, J. and {Bouwhuis}, M. and {Dodson}, R. and {Ekers}, R.~D. and {Farah}, W. and {Flynn}, C. and {James}, C.~W. and {Kerr}, M. and {Lenc}, E. and {Mahony}, E.~K. and {O'Meara}, J. and {Os{\l}owski}, S. and {Qiu}, H. and {Treu}, T. and {U}, V. and {Bateman}, T.~J. and {Bock}, D.~C.-J. and {Bolton}, R.~J. and {Brown}, A. and {Bunton}, J.~D. and {Chippendale}, A.~P. and {Cooray}, F.~R. and {Cornwell}, T. and {Gupta}, N. and {Hayman}, D.~B. and {Kesteven}, M. and {Koribalski}, B.~S. and {MacLeod}, A. and {McClure-Griffiths}, N.~M. and {Neuhold}, S. and {Norris}, R.~P. and {Pilawa}, M.~A. and {Qiao}, R.-Y. and {Reynolds}, J. and {Roxby}, D.~N. and {Shimwell}, T.~W. and {Voronkov}, M.~A. and {Wilson}, C.~D.},
        title = "{A single fast radio burst localized to a massive galaxy at cosmological distance}",
      journal = {Science},
         year = 2019,
        month = aug,
       volume = {365},
       number = {6453},
        pages = {565-570},
          doi = {10.1126/science.aaw5903},
archivePrefix = {arXiv},
       eprint = {1906.11476},
 primaryClass = {astro-ph.HE},
       adsurl = {https://ui.adsabs.harvard.edu/abs/2019Sci...365..565B}
}

@ARTICLE{2025Natur.647..603C,
       author = {{Callingham}, J.~R. and {Tasse}, C. and {Keers}, R. and {Kavanagh}, R.~D. and {Vedantham}, H.~K. and {Zarka}, P. and {Bellotti}, S. and {Cristofari}, P.~I. and {Bloot}, S. and {Konijn}, D.~C. and {Hardcastle}, M.~J. and {Lamy}, L. and {Pass}, E.~K. and {Pope}, B.~J.~S. and {Reid}, H. and {R{\"o}ttgering}, H.~J.~A. and {Shimwell}, T.~W. and {Zucca}, P.},
        title = "{Radio burst from a stellar coronal mass ejection}",
      journal = {\nat},
         year = 2025,
        month = nov,
       volume = {647},
       number = {8090},
        pages = {603-607},
          doi = {10.1038/s41586-025-09715-3},
archivePrefix = {arXiv},
       eprint = {2511.09289},
 primaryClass = {astro-ph.SR},
       adsurl = {https://ui.adsabs.harvard.edu/abs/2025Natur.647..603C}
}

@article{CARILLI2004979,
title = {Motivation, key science projects, standards and assumptions},
journal = {New Astronomy Reviews},
volume = {48},
number = {11},
pages = {979-984},
year = {2004},
note = {Science with the Square Kilometre Array},
issn = {1387-6473},
doi = {https://doi.org/10.1016/j.newar.2004.09.001},
url = {https://www.sciencedirect.com/science/article/pii/S1387647304000880},
author = {C.L. Carilli and S. Rawlings}
}

@ARTICLE{2018arXiv180709080C,
       author = {{Cassano}, R. and {Fender}, R. and {Ferrari}, C. and {Merloni}, A. and {Akahori}, T. and {Akamatsu}, H. and {Ascasibar}, Y. and {Ballantyne}, D. and {Brunetti}, G. and {Corbelli}, E. and {Croston}, J. and {Donnarumma}, I. and {Ettori}, S. and {Ferdman}, R. and {Feretti}, L. and {Forbrich}, J. and {Gheller}, C. and {Ghirlanda}, G. and {Govoni}, F. and {Ingallinera}, A. and {Johnston-Hollitt}, M. and {Markevitch}, M. and {Mesinger}, A. and {Moss}, V. and {Nicastro}, F. and {Padovani}, P. and {Panessa}, F. and {Piro}, L. and {Ponti}, G. and {Pratt}, G. and {Rossi}, E.~M. and {Sadler}, E. and {Sasaki}, M. and {Soria}, R. and {Stevens}, I. and {van Weeren}, R. and {Vazza}, F. and {Webb}, N.},
        title = "{SKA-Athena Synergy White Paper}",
      journal = {arXiv e-prints},
         year = 2018,
        month = jul,
          eid = {arXiv:1807.09080},
        pages = {arXiv:1807.09080},
          doi = {10.48550/arXiv.1807.09080},
archivePrefix = {arXiv},
       eprint = {1807.09080},
 primaryClass = {astro-ph.HE},
       adsurl = {https://ui.adsabs.harvard.edu/abs/2018arXiv180709080C}
}

@ARTICLE{2009arXiv0912.0201L,
       author = {{LSST Science Collaboration} and {Abell}, Paul A. and {Allison}, Julius and {Anderson}, Scott F. and {Andrew}, John R. and {Angel}, J. Roger P. and {Armus}, Lee and {Arnett}, David and {Asztalos}, S.~J. and {Axelrod}, Tim S. and {Bailey}, Stephen and {Ballantyne}, D.~R. and {Bankert}, Justin R. and {Barkhouse}, Wayne A. and {Barr}, Jeffrey D. and {Barrientos}, L. Felipe and {Barth}, Aaron J. and {Bartlett}, James G. and {Becker}, Andrew C. and {Becla}, Jacek and {Beers}, Timothy C. and {Bernstein}, Joseph P. and {Biswas}, Rahul and {Blanton}, Michael R. and {Bloom}, Joshua S. and {Bochanski}, John J. and {Boeshaar}, Pat and {Borne}, Kirk D. and {Bradac}, Marusa and {Brandt}, W.~N. and {Bridge}, Carrie R. and {Brown}, Michael E. and {Brunner}, Robert J. and {Bullock}, James S. and {Burgasser}, Adam J. and {Burge}, James H. and {Burke}, David L. and {Cargile}, Phillip A. and {Chandrasekharan}, Srinivasan and {Chartas}, George and {Chesley}, Steven R. and {Chu}, You-Hua and {Cinabro}, David and {Claire}, Mark W. and {Claver}, Charles F. and {Clowe}, Douglas and {Connolly}, A.~J. and {Cook}, Kem H. and {Cooke}, Jeff and {Cooray}, Asantha and {Covey}, Kevin R. and {Culliton}, Christopher S. and {de Jong}, Roelof and {de Vries}, Willem H. and {Debattista}, Victor P. and {Delgado}, Francisco and {Dell'Antonio}, Ian P. and {Dhital}, Saurav and {Di Stefano}, Rosanne and {Dickinson}, Mark and {Dilday}, Benjamin and {Djorgovski}, S.~G. and {Dobler}, Gregory and {Donalek}, Ciro and {Dubois-Felsmann}, Gregory and {Durech}, Josef and {Eliasdottir}, Ardis and {Eracleous}, Michael and {Eyer}, Laurent and {Falco}, Emilio E. and {Fan}, Xiaohui and {Fassnacht}, Christopher D. and {Ferguson}, Harry C. and {Fernandez}, Yanga R. and {Fields}, Brian D. and {Finkbeiner}, Douglas and {Figueroa}, Eduardo E. and {Fox}, Derek B. and {Francke}, Harold and {Frank}, James S. and {Frieman}, Josh and {Fromenteau}, Sebastien and {Furqan}, Muhammad and {Galaz}, Gaspar and {Gal-Yam}, A. and {Garnavich}, Peter and {Gawiser}, Eric and {Geary}, John and {Gee}, Perry and {Gibson}, Robert R. and {Gilmore}, Kirk and {Grace}, Emily A. and {Green}, Richard F. and {Gressler}, William J. and {Grillmair}, Carl J. and {Habib}, Salman and {Haggerty}, J.~S. and {Hamuy}, Mario and {Harris}, Alan W. and {Hawley}, Suzanne L. and {Heavens}, Alan F. and {Hebb}, Leslie and {Henry}, Todd J. and {Hileman}, Edward and {Hilton}, Eric J. and {Hoadley}, Keri and {Holberg}, J.~B. and {Holman}, Matt J. and {Howell}, Steve B. and {Infante}, Leopoldo and {Ivezic}, Zeljko and {Jacoby}, Suzanne H. and {Jain}, Bhuvnesh and {R} and {Jedicke} and {Jee}, M. James and {Garrett Jernigan}, J. and {Jha}, Saurabh W. and {Johnston}, Kathryn V. and {Jones}, R. Lynne and {Juric}, Mario and {Kaasalainen}, Mikko and {Styliani} and {Kafka} and {Kahn}, Steven M. and {Kaib}, Nathan A. and {Kalirai}, Jason and {Kantor}, Jeff and {Kasliwal}, Mansi M. and {Keeton}, Charles R. and {Kessler}, Richard and {Knezevic}, Zoran and {Kowalski}, Adam and {Krabbendam}, Victor L. and {Krughoff}, K. Simon and {Kulkarni}, Shrinivas and {Kuhlman}, Stephen and {Lacy}, Mark and {Lepine}, Sebastien and {Liang}, Ming and {Lien}, Amy and {Lira}, Paulina and {Long}, Knox S. and {Lorenz}, Suzanne and {Lotz}, Jennifer M. and {Lupton}, R.~H. and {Lutz}, Julie and {Macri}, Lucas M. and {Mahabal}, Ashish A. and {Mandelbaum}, Rachel and {Marshall}, Phil and {May}, Morgan and {McGehee}, Peregrine M. and {Meadows}, Brian T. and {Meert}, Alan and {Milani}, Andrea and {Miller}, Christopher J. and {Miller}, Michelle and {Mills}, David and {Minniti}, Dante and {Monet}, David and {Mukadam}, Anjum S. and {Nakar}, Ehud and {Neill}, Douglas R. and {Newman}, Jeffrey A. and {Nikolaev}, Sergei and {Nordby}, Martin and {O'Connor}, Paul and {Oguri}, Masamune and {Oliver}, John and {Olivier}, Scot S. and {Olsen}, Julia K. and {Olsen}, Knut and {Olszewski}, Edward W. and {Oluseyi}, Hakeem and {Padilla}, Nelson D. and {Parker}, Alex and {Pepper}, Joshua and {Peterson}, John R. and {Petry}, Catherine and {Pinto}, Philip A. and {Pizagno}, James L. and {Popescu}, Bogdan and {Prsa}, Andrej and {Radcka}, Veljko and {Raddick}, M. Jordan and {Rasmussen}, Andrew and {Rau}, Arne and {Rho}, Jeonghee and {Rhoads}, James E. and {Richards}, Gordon T. and {Ridgway}, Stephen T. and {Robertson}, Brant E. and {Roskar}, Rok and {Saha}, Abhijit and {Sarajedini}, Ata and {Scannapieco}, Evan and {Schalk}, Terry and {Schindler}, Rafe and {Schmidt}, Samuel},
        title = "{LSST Science Book, Version 2.0}",
      journal = {arXiv e-prints},
         year = 2009,
        month = dec,
          eid = {arXiv:0912.0201},
        pages = {arXiv:0912.0201},
          doi = {10.48550/arXiv.0912.0201},
archivePrefix = {arXiv},
       eprint = {0912.0201},
 primaryClass = {astro-ph.IM},
       adsurl = {https://ui.adsabs.harvard.edu/abs/2009arXiv0912.0201L}
}

@ARTICLE{2018ApJ...858L..15D,
       author = {{Dobie}, Dougal and {Kaplan}, David L. and {Murphy}, Tara and {Lenc}, Emil and {Mooley}, Kunal P. and {Lynch}, Christene and {Corsi}, Alessandra and {Frail}, Dale and {Kasliwal}, Mansi and {Hallinan}, Gregg},
        title = "{A Turnover in the Radio Light Curve of GW170817}",
      journal = {\apjl},
         year = 2018,
        month = may,
       volume = {858},
       number = {2},
          eid = {L15},
        pages = {L15},
          doi = {10.3847/2041-8213/aac105},
archivePrefix = {arXiv},
       eprint = {1803.06853},
 primaryClass = {astro-ph.HE},
       adsurl = {https://ui.adsabs.harvard.edu/abs/2018ApJ...858L..15D}
}

@ARTICLE{2017Sci...358.1579H,
       author = {{Hallinan}, G. and {Corsi}, A. and {Mooley}, K.~P. and {Hotokezaka}, K. and {Nakar}, E. and {Kasliwal}, M.~M. and {Kaplan}, D.~L. and {Frail}, D.~A. and {Myers}, S.~T. and {Murphy}, T. and {De}, K. and {Dobie}, D. and {Allison}, J.~R. and {Bannister}, K.~W. and {Bhalerao}, V. and {Chandra}, P. and {Clarke}, T.~E. and {Giacintucci}, S. and {Ho}, A.~Y.~Q. and {Horesh}, A. and {Kassim}, N.~E. and {Kulkarni}, S.~R. and {Lenc}, E. and {Lockman}, F.~J. and {Lynch}, C. and {Nichols}, D. and {Nissanke}, S. and {Palliyaguru}, N. and {Peters}, W.~M. and {Piran}, T. and {Rana}, J. and {Sadler}, E.~M. and {Singer}, L.~P.},
        title = "{A radio counterpart to a neutron star merger}",
      journal = {Science},
         year = 2017,
        month = dec,
       volume = {358},
       number = {6370},
        pages = {1579-1583},
          doi = {10.1126/science.aap9855},
archivePrefix = {arXiv},
       eprint = {1710.05435},
 primaryClass = {astro-ph.HE},
       adsurl = {https://ui.adsabs.harvard.edu/abs/2017Sci...358.1579H}
}

@ARTICLE{2019ApJ...871...73H,
       author = {{Ho}, Anna Y.~Q. and {Phinney}, E. Sterl and {Ravi}, Vikram and {Kulkarni}, S.~R. and {Petitpas}, Glen and {Emonts}, Bjorn and {Bhalerao}, V. and {Blundell}, Ray and {Cenko}, S. Bradley and {Dobie}, Dougal and {Howie}, Ryan and {Kamraj}, Nikita and {Kasliwal}, Mansi M. and {Murphy}, Tara and {Perley}, Daniel A. and {Sridharan}, T.~K. and {Yoon}, Ilsang},
        title = "{AT2018cow: A Luminous Millimeter Transient}",
      journal = {\apj},
         year = 2019,
        month = jan,
       volume = {871},
       number = {1},
          eid = {73},
        pages = {73},
          doi = {10.3847/1538-4357/aaf473},
archivePrefix = {arXiv},
       eprint = {1810.10880},
 primaryClass = {astro-ph.HE},
       adsurl = {https://ui.adsabs.harvard.edu/abs/2019ApJ...871...73H}
}

@INPROCEEDINGS{2016mks..confE..13F,
       author = {{Fender}, R. and {Woudt}, P.~A. and {Corbel}, S. and {Coriat}, M. and {Daigne}, F. and {Falcke}, H. and {Girard}, J. and {Heywood}, I. and {Horesh}, A. and {Horrell}, J. and {Jonker}, P.~G. and {Joseph}, T. and {Kamble}, A. and {Knigge}, C. and {K{\"o}rding}, E. and {Kotze}, M. and {Kouveliotou}, C. and {Lynch}, C. and {Maccarone}, T. and {Meintjes}, P. and {Migliari}, S. and {Murphy}, T. and {Nagayama}, T. and {Nelemans}, G. and {Nicholson}, G. and {O'Brien}, T. and {Oodendaal}, A. and {Oozeer}, N. and {Osborne}, J. and {P{\'e}rez-Torres}, M. and {Ratcliffe}, S. and {Ribeiro}, V.~A.~R.~M. and {Rol}, E. and {Rushton}, A. and {Scaife}, A. and {Schurch}, M. and {Sivakoff}, G. and {Staley}, T. and {Steeghs}, D. and {Stewart}, I. and {Swinbank}, J.~D. and {Vergani}, S. and {Warner}, B. and {Wiersema}, K. and {Armstrong}, R. and {Groot}, P. and {McBride}, V. and {Miller-Jones}, J.~C.~A. and {Mooley}, K. and {Stappers}, B. and {Wijers}, R.~A.~M.~J. and {Bietenholz}, M. and {Blyth}, S. and {B{\"o}ttcher}, M. and {Buckley}, D. and {Charles}, P. and {Chomiuk}, L. and {Coppejans}, D. and {de Blok}, W.~J.~G. and {van der Heyden}, K. and {van der Horst}, A. and {van Soelen}, B.},
        title = "{ThunderKAT: The MeerKAT Large Survey Project for Image-Plane Radio Transients}",
    booktitle = {MeerKAT Science: On the Pathway to the SKA},
         year = 2016,
        month = jan,
          eid = {13},
        pages = {13},
          doi = {10.22323/1.277.0013},
archivePrefix = {arXiv},
       eprint = {1711.04132},
 primaryClass = {astro-ph.HE},
       adsurl = {https://ui.adsabs.harvard.edu/abs/2016mks..confE..13F}
}

@INPROCEEDINGS{2015aska.confE..51F,
       author = {{Fender}, R. and {Stewart}, A. and {Macquart}, J.~P. and {Donnarumma}, I. and {Murphy}, T. and {Deller}, A. and {Paragi}, Z. and {Chatterjee}, S.},
        title = "{The Transient Universe with the Square Kilometre Array}",
    booktitle = {Advancing Astrophysics with the Square Kilometre Array (AASKA14)},
         year = 2015,
        month = apr,
          eid = {51},
        pages = {51},
          doi = {10.22323/1.215.0051},
archivePrefix = {arXiv},
       eprint = {1507.00729},
 primaryClass = {astro-ph.HE},
       adsurl = {https://ui.adsabs.harvard.edu/abs/2015aska.confE..51F}
}

@INPROCEEDINGS{2015aska.confE..55M,
       author = {{Macquart}, J.~P. and {Keane}, E. and {Grainge}, K. and {McQuinn}, M. and {Fender}, R. and {Hessels}, J. and {Deller}, A. and {Bhat}, R. and {Breton}, R. and {Chatterjee}, S. and {Law}, C. and {Lorimer}, D. and {Ofek}, E.~O. and {Pietka}, M. and {Spitler}, L. and {Stappers}, B. and {Trott}, C.},
        title = "{Fast Transients at Cosmological Distances with the SKA}",
    booktitle = {Advancing Astrophysics with the Square Kilometre Array (AASKA14)},
         year = 2015,
        month = apr,
          eid = {55},
        pages = {55},
          doi = {10.22323/1.215.0055},
archivePrefix = {arXiv},
       eprint = {1501.07535},
 primaryClass = {astro-ph.HE},
       adsurl = {https://ui.adsabs.harvard.edu/abs/2015aska.confE..55M}
}

@ARTICLE{2015MNRAS.446.3687P,
       author = {{Pietka}, M. and {Fender}, R.~P. and {Keane}, E.~F.},
        title = "{The variability time-scales and brightness temperatures of radio flares from stars to supermassive black holes}",
      journal = {\mnras},
         year = 2015,
        month = feb,
       volume = {446},
       number = {4},
        pages = {3687-3696},
          doi = {10.1093/mnras/stu2335},
archivePrefix = {arXiv},
       eprint = {1411.1067},
 primaryClass = {astro-ph.HE},
       adsurl = {https://ui.adsabs.harvard.edu/abs/2015MNRAS.446.3687P}
}

@ARTICLE{2004NewAR..48.1459C,
       author = {{Cordes}, James M. and {Lazio}, T. Joseph W. and {McLaughlin}, M.~A.},
        title = "{The dynamic radio sky}",
      journal = {\nar},
         year = 2004,
        month = dec,
       volume = {48},
       number = {11-12},
        pages = {1459-1472},
          doi = {10.1016/j.newar.2004.09.038},
archivePrefix = {arXiv},
       eprint = {astro-ph/0410045},
 primaryClass = {astro-ph},
       adsurl = {https://ui.adsabs.harvard.edu/abs/2004NewAR..48.1459C}
}

@BOOK{1984codi.book.....H,
       author = {{Harwit}, Martin},
        title = "{Cosmic Discovery}",
         year = 1984,
       adsurl = {https://ui.adsabs.harvard.edu/abs/1984codi.book.....H}
}

@INPROCEEDINGS{2015aska.confE..65W,
       author = {{Wilkinson}, P.},
        title = "{The SKA and the Unknown Unknowns}",
    booktitle = {Advancing Astrophysics with the Square Kilometre Array (AASKA14)},
         year = 2015,
        month = apr,
          eid = {65},
        pages = {65},
          doi = {10.22323/1.215.0065},
archivePrefix = {arXiv},
       eprint = {1506.08540},
 primaryClass = {astro-ph.IM},
       adsurl = {https://ui.adsabs.harvard.edu/abs/2015aska.confE..65W}
}

@ARTICLE{2004NewAR..48.1551W,
       author = {{Wilkinson}, P.~N. and {Kellermann}, K.~I. and {Ekers}, R.~D. and {Cordes}, J.~M. and {W. Lazio}, T. Joseph},
        title = "{The exploration of the unknown}",
      journal = {\nar},
         year = 2004,
        month = dec,
       volume = {48},
       number = {11-12},
        pages = {1551-1563},
          doi = {10.1016/j.newar.2004.09.036},
archivePrefix = {arXiv},
       eprint = {astro-ph/0410225},
 primaryClass = {astro-ph},
       adsurl = {https://ui.adsabs.harvard.edu/abs/2004NewAR..48.1551W}
}

@INPROCEEDINGS{2014aska,
       author = {{Braun}, R. and {Bourke}, T. and {Green}, J.~A. and {Keane}, E. and {Wagg}, J.},
        title = "{Advancing Astrophysics with the Square Kilometre Array}",
    booktitle = {Advancing Astrophysics with the Square Kilometre Array (AASKA14)},
         year = 2015,
        month = apr,
          eid = {174},
        pages = {174},
          doi = {10.22323/1.215.0174},
       adsurl = {https://ui.adsabs.harvard.edu/abs/2015aska.confE.174B}
}

@BOOK{2024skabook,
       author = {{Schilizzi}, Richard T. and {Ekers}, Ronald D. and {Dewdney}, Peter E. and {Crosby}, Philip},
        title = "{The Square Kilometre Array. A Science Mega-Project in the Making, 1990-2012}",
         year = 2024,
          doi = {10.1007/978-3-031-51374-9},
       adsurl = {https://ui.adsabs.harvard.edu/abs/2024ska..book.....S}
}

@ARTICLE{2006ApJ...639..331G,
       author = {{Gal-Yam}, Avishay and {Ofek}, Eran O. and {Poznanski}, Dovi and {Levinson}, Amir and {Waxman}, Eli and {Frail}, Dale A. and {Soderberg}, Alicia M. and {Nakar}, Ehud and {Li}, Weidong and {Filippenko}, Alexei V.},
        title = "{Radio and Optical Follow-up Observations of a Uniform Radio Transient Search: Implications for Gamma-Ray Bursts and Supernovae}",
      journal = {\apj},
         year = 2006,
        month = mar,
       volume = {639},
       number = {1},
        pages = {331-339},
          doi = {10.1086/499157},
archivePrefix = {arXiv},
       eprint = {astro-ph/0508629},
 primaryClass = {astro-ph},
       adsurl = {https://ui.adsabs.harvard.edu/abs/2006ApJ...639..331G}
}

@ARTICLE{2024AJ....167....5L,
       author = {{Luhman}, K.~L. and {Tremblin}, P. and {Alves de Oliveira}, C. and {Birkmann}, S.~M. and {Baraffe}, I. and {Chabrier}, G. and {Manjavacas}, E. and {Parker}, R.~J. and {Valenti}, J.},
        title = "{JWST/NIRSpec Observations of the Coldest Known Brown Dwarf}",
      journal = {\aj},
         year = 2024,
        month = jan,
       volume = {167},
       number = {1},
          eid = {5},
        pages = {5},
          doi = {10.3847/1538-3881/ad0b72},
archivePrefix = {arXiv},
       eprint = {2311.17316},
 primaryClass = {astro-ph.SR},
       adsurl = {https://ui.adsabs.harvard.edu/abs/2024AJ....167....5L}
}

@ARTICLE{2010A&A...522A..13R,
       author = {{Reiners}, A. and {Christensen}, U.~R.},
        title = "{A magnetic field evolution scenario for brown dwarfs and giant planets}",
      journal = {\aap},
         year = 2010,
        month = nov,
       volume = {522},
          eid = {A13},
        pages = {A13},
          doi = {10.1051/0004-6361/201014251},
archivePrefix = {arXiv},
       eprint = {1007.1514},
 primaryClass = {astro-ph.EP},
       adsurl = {https://ui.adsabs.harvard.edu/abs/2010A&A...522A..13R}
}

@ARTICLE{2025A&A...699A.341B,
       author = {{Bloot}, S. and {Vedantham}, H.~K. and {Bassa}, C.~G. and {Callingham}, J.~R. and {Best}, W.~M.~J. and {Liu}, M.~C. and {Magnier}, E.~A. and {Shimwell}, T.~W. and {Dupuy}, T.~J.},
        title = "{Strongly polarised radio pulses from a new white-dwarf-hosting long-period transient}",
      journal = {\aap},
         year = 2025,
        month = jul,
       volume = {699},
          eid = {A341},
        pages = {A341},
          doi = {10.1051/0004-6361/202555131},
archivePrefix = {arXiv},
       eprint = {2507.05078},
 primaryClass = {astro-ph.HE},
       adsurl = {https://ui.adsabs.harvard.edu/abs/2025A&A...699A.341B}
}

@Article{Connor2018,
  author    = {Connor, Liam and van Leeuwen, Joeri},
  journal   = {The Astronomical Journal},
  title     = {Applying Deep Learning to Fast Radio Burst Classification},
  year      = {2018},
  issn      = {1538-3881},
  month     = nov,
  number    = {6},
  pages     = {256},
  volume    = {156},
  doi       = {10.3847/1538-3881/aae649},
  publisher = {American Astronomical Society},
}

@ARTICLE{2020MNRAS.497.1661A,
       author = {{Agarwal}, Devansh and {Aggarwal}, Kshitij and {Burke-Spolaor}, Sarah and {Lorimer}, Duncan R. and {Garver-Daniels}, Nathaniel},
        title = "{FETCH: A deep-learning based classifier for fast transient classification}",
      journal = {Monthly Notices of the Royal Astronomical Society},
         year = 2020,
        month = sep,
       volume = {497},
       number = {2},
        pages = {1661-1674},
          doi = {10.1093/mnras/staa1856},
archivePrefix = {arXiv},
       eprint = {1902.06343},
 primaryClass = {astro-ph.IM},
       adsurl = {https://ui.adsabs.harvard.edu/abs/2020MNRAS.497.1661A}
}

@ARTICLE{Shu2025_FXT_mildly_rel,
       author = {{Shu}, Xinwen and {Yang}, Lei and {Yang}, Haonan and {Xu}, Fan and {Chen}, Jin-Hong and {Eyles-Ferris}, Rob A.~J. and {Dai}, Lixin and {Yu}, Yunwei and {Shen}, Rong-Feng and {Sun}, Luming and {Ding}, Hucheng and {Zheng}, WeiKang and {Jiang}, Ning and {Li}, Wenxiong and {Sun}, Ning-Chen and {Xu}, Dong and {Zhang}, Zhumao and {Jin}, Chichuan and {Rau}, Arne and {Wang}, Tinggui and {Wu}, Xue-feng and {Yuan}, Weimin and {Zhang}, Bing and {Nandra}, Kirpal and {Filippenko}, Alexei V. and {Poidevin}, Fr{\'e}d{\'e}rick and {Soria}, Roberto and {Kumar}, Amit and {Aguado}, David S. and {An}, Fangxia and {An}, Tao and {An}, Jie and {Andrews}, Moira and {Anutarawiramkul}, Rungrit and {Baldini}, Pietro and {Brink}, Thomas G. and {Butpan}, Pathompong and {Cai}, Zhiming and {Castro-Tirado}, Alberto J. and {Cheng}, Huaqing and {Cui}, Weiwei and {Farah}, Joseph and {Fu}, Shaoyu and {Fynbo}, Johan P.~U. and {Gao}, Xing and {Han}, Dawei and {Han}, Xuhui and {Howell}, D. Andrew and {Hu}, Jingwei and {Jiang}, Shuaiqing and {Kumar}, Brajesh and {Lei}, Weihua and {Li}, Dongyue and {Li}, Chengkui and {Liu}, Huaqiu and {Liu}, Xing and {Liu}, Yuan and {Liu}, Xiaowei and {L{\'o}pez-Oramas}, Alicia and {L{\'o}pez Fern{\'a}ndez-Nespral}, David and {Maund}, Justyn R. and {McCully}, Curtis and {Niu}, Zexi and {Newsome}, Megan and {O'Brien}, Paul and {Pan}, Haiwu and {Pan}, Yu and {Padilla Gonzalez}, Estefania and {P{\'e}rez-Fournon}, Ismael and {Silima}, Walter and {Sun}, Hui and {Sun}, Shengli and {Sun}, Xiaojin and {Terreran}, Giacomo and {Tinyanont}, Samaporn and {Wang}, Junxian and {Wang}, Yanan and {Wang}, Yun and {Wiersema}, Klaas and {Xu}, Yunfei and {Xue}, Yongquan and {Yang}, Yi and {Zhang}, Fabao and {Zhang}, Juan and {Zhang}, Pinpin and {Zhang}, Wenda and {Zhang}, Yonghe and {Zhao}, Haisheng and {Zhu}, Zipei and {Xin}, Liping and {Yao}, Zhuheng and {Cordier}, Bertrand and {Wei}, Jianyan and {Qiu}, Yulei and {Daigne}, Fr{\'e}d{\'e}ric},
        title = "{EP241021a: A Months-duration X-Ray Transient with Luminous Optical and Radio Emission}",
      journal = {\apjl},
         year = 2025,
        month = sep,
       volume = {990},
       number = {1},
          eid = {L29},
        pages = {L29},
          doi = {10.3847/2041-8213/adf4cd},
archivePrefix = {arXiv},
       eprint = {2505.07665},
 primaryClass = {astro-ph.HE},
       adsurl = {https://ui.adsabs.harvard.edu/abs/2025ApJ...990L..29S}
}

@ARTICLE{hessels19,
       author = {{Hessels}, J.~W.~T. and {Spitler}, L.~G. and {Seymour}, A.~D. and {Cordes}, J.~M. and {Michilli}, D. and {Lynch}, R.~S. and {Gourdji}, K. and {Archibald}, A.~M. and {Bassa}, C.~G. and {Bower}, G.~C. and {Chatterjee}, S. and {Connor}, L. and {Crawford}, F. and {Deneva}, J.~S. and {Gajjar}, V. and {Kaspi}, V.~M. and {Keimpema}, A. and {Law}, C.~J. and {Marcote}, B. and {McLaughlin}, M.~A. and {Paragi}, Z. and {Petroff}, E. and {Ransom}, S.~M. and {Scholz}, P. and {Stappers}, B.~W. and {Tendulkar}, S.~P.},
        title = "{FRB 121102 Bursts Show Complex Time-Frequency Structure}",
      journal = {\apjl},
         year = 2019,
        month = may,
       volume = {876},
       number = {2},
          eid = {L23},
        pages = {L23},
          doi = {10.3847/2041-8213/ab13ae},
archivePrefix = {arXiv},
       eprint = {1811.10748},
 primaryClass = {astro-ph.HE},
       adsurl = {https://ui.adsabs.harvard.edu/abs/2019ApJ...876L..23H}
}

@ARTICLE{platts_2019_frbmodels,
       author = {{Platts}, E. and {Weltman}, A. and {Walters}, A. and {Tendulkar}, S.~P. and {Gordin}, J.~E.~B. and {Kandhai}, S.},
        title = "{A living theory catalogue for fast radio bursts}",
      journal = {\physrep},
         year = 2019,
        month = aug,
       volume = {821},
        pages = {1-27},
          doi = {10.1016/j.physrep.2019.06.003},
archivePrefix = {arXiv},
       eprint = {1810.05836},
 primaryClass = {astro-ph.HE},
       adsurl = {https://ui.adsabs.harvard.edu/abs/2019PhR...821....1P}
}

@ARTICLE{pm_2023_frbquasi,
       author = {{Pastor-Marazuela}, In{\'e}s and {van Leeuwen}, Joeri and {Bilous}, Anna and {Connor}, Liam and {Maan}, Yogesh and {Oostrum}, Leon and {Petroff}, Emily and {Straal}, Samayra and {Vohl}, Dany and {Adams}, Elizabeth A.~K. and {Adebahr}, Bj{\"o}rn and {Attema}, Jisk and {Boersma}, Oliver M. and {van den Brink}, R. and {van Cappellen}, W.~A. and {Coolen}, Arthur H.~W.~M. and {Damstra}, Sieds and {D{\'e}nes}, Helga and {Hess}, Kelley M. and {van der Hulst}, J.~M. and {Hut}, Boudewijn and {Kutkin}, Alexander and {Marcel Loose}, G. and {Lucero}, Danielle M. and {Mika}, {\'A}gnes and {Moss}, Vanessa A. and {Mulder}, Henk and {Norden}, Menno J. and {Oosterloo}, Tom A. and {Rajwade}, Kaustubh and {van der Schuur}, Daniel and {Sclocco}, Alessio and {Smits}, R. and {Ziemke}, Jacob},
        title = "{A fast radio burst with submillisecond quasi-periodic structure}",
      journal = {\aap},
         year = 2023,
        month = oct,
       volume = {678},
          eid = {A149},
        pages = {A149},
          doi = {10.1051/0004-6361/202243339},
archivePrefix = {arXiv},
       eprint = {2202.08002},
 primaryClass = {astro-ph.HE},
       adsurl = {https://ui.adsabs.harvard.edu/abs/2023A&A...678A.149P}
}

@ARTICLE{caleb_2022_76spsr,
       author = {{Caleb}, Manisha and {Heywood}, Ian and {Rajwade}, Kaustubh and {Malenta}, Mateusz and {Stappers}, Benjamin Willem and {Barr}, Ewan and {Chen}, Weiwei and {Morello}, Vincent and {Sanidas}, Sotiris and {van den Eijnden}, Jakob and {Kramer}, Michael and {Buckley}, David and {Brink}, Jaco and {Motta}, Sara Elisa and {Woudt}, Patrick and {Weltevrede}, Patrick and {Jankowski}, Fabian and {Surnis}, Mayuresh and {Buchner}, Sarah and {Bezuidenhout}, Mechiel Christiaan and {Driessen}, Laura Nicole and {Fender}, Rob},
        title = "{Discovery of a radio-emitting neutron star with an ultra-long spin period of 76 s}",
      journal = {Nature Astronomy},
         year = 2022,
        month = may,
       volume = {6},
        pages = {828-836},
          doi = {10.1038/s41550-022-01688-x},
archivePrefix = {arXiv},
       eprint = {2206.01346},
 primaryClass = {astro-ph.HE},
       adsurl = {https://ui.adsabs.harvard.edu/abs/2022NatAs...6..828C}
}

@ARTICLE{kramer_2024_quasiperiod,
       author = {{Kramer}, Michael and {Liu}, Kuo and {Desvignes}, Gregory and {Karuppusamy}, Ramesh and {Stappers}, Ben W.},
        title = "{Quasi-periodic sub-pulse structure as a unifying feature for radio-emitting neutron stars}",
      journal = {Nature Astronomy},
         year = 2024,
        month = feb,
       volume = {8},
        pages = {230-240},
          doi = {10.1038/s41550-023-02125-3},
archivePrefix = {arXiv},
       eprint = {2311.13762},
 primaryClass = {astro-ph.HE},
       adsurl = {https://ui.adsabs.harvard.edu/abs/2024NatAs...8..230K}
}

@ARTICLE{weiler_2002,
       author = {{Weiler}, Kurt W. and {Panagia}, Nino and {Montes}, Marcos J. and {Sramek}, Richard A.},
        title = "{Radio Emission from Supernovae and Gamma-Ray Bursters}",
      journal = {\araa},
         year = 2002,
        month = jan,
       volume = {40},
        pages = {387-438},
          doi = {10.1146/annurev.astro.40.060401.093744},
       adsurl = {https://ui.adsabs.harvard.edu/abs/2002ARA&A..40..387W}
}

@ARTICLE{weiler_2007_sn1993j,
       author = {{Weiler}, Kurt W. and {Williams}, Christopher L. and {Panagia}, Nino and {Stockdale}, Christopher J. and {Kelley}, Matthew T. and {Sramek}, Richard A. and {Van Dyk}, Schuyler D. and {Marcaide}, J.~M.},
        title = "{Long-Term Radio Monitoring of SN 1993J}",
      journal = {\apj},
         year = 2007,
        month = dec,
       volume = {671},
       number = {2},
        pages = {1959-1980},
          doi = {10.1086/523258},
archivePrefix = {arXiv},
       eprint = {0709.1136},
 primaryClass = {astro-ph},
       adsurl = {https://ui.adsabs.harvard.edu/abs/2007ApJ...671.1959W}
}

@ARTICLE{martividal_2011_sn1993j2,
       author = {{Mart{\'\i}-Vidal}, I. and {Marcaide}, J.~M. and {Alberdi}, A. and {Guirado}, J.~C. and {P{\'e}rez-Torres}, M.~A. and {Ros}, E.},
        title = "{Radio emission of SN1993J: the complete picture. II. Simultaneous fit of expansion and radio light curves}",
      journal = {\aap},
         year = 2011,
        month = feb,
       volume = {526},
          eid = {A143},
        pages = {A143},
          doi = {10.1051/0004-6361/201014517},
archivePrefix = {arXiv},
       eprint = {1007.1224},
 primaryClass = {astro-ph.HE},
       adsurl = {https://ui.adsabs.harvard.edu/abs/2011A&A...526A.143M}
}

@ARTICLE{granot_2014,
       author = {{Granot}, Jonathan and {van der Horst}, Alexander J.},
        title = "{Gamma-Ray Burst Jets and their Radio Observations}",
      journal = {\pasa},
         year = 2014,
        month = feb,
       volume = {31},
          eid = {e008},
        pages = {e008},
          doi = {10.1017/pasa.2013.44},
       adsurl = {https://ui.adsabs.harvard.edu/abs/2014PASA...31....8G}
}

@ARTICLE{pulsar,
       author = {{Hewish}, A. and {Bell}, S.~J. and {Pilkington}, J.~D.~H. and {Scott}, P.~F. and {Collins}, R.~A.},
        title = "{Observation of a Rapidly Pulsating Radio Source}",
      journal = {\nat},
         year = 1968,
        month = feb,
       volume = {217},
       number = {5130},
        pages = {709-713},
          doi = {10.1038/217709a0},
       adsurl = {https://ui.adsabs.harvard.edu/abs/1968Natur.217..709H}
}

@ARTICLE{kumarsub,
       author = {{Kumar}, P. and {Shannon}, R.~M. and {Flynn}, C. and {Os{\l}owski}, S. and {Bhandari}, S. and {Day}, C.~K. and {Deller}, A.~T. and {Farah}, W. and {Kaczmarek}, J.~F. and {Kerr}, M. and {Phillips}, C. and {Price}, D.~C. and {Qiu}, H. and {Thyagarajan}, N.},
        title = "{Extremely band-limited repetition from a fast radio burst source}",
      journal = {\mnras},
         year = 2021,
        month = jan,
       volume = {500},
       number = {2},
        pages = {2525-2531},
          doi = {10.1093/mnras/staa3436},
archivePrefix = {arXiv},
       eprint = {2009.01214},
 primaryClass = {astro-ph.HE},
       adsurl = {https://ui.adsabs.harvard.edu/abs/2021MNRAS.500.2525K}
}

\end{document}